\documentclass[aps,prl,twocolumn,groupedaddress,amsmath,amssymb,nofootinbib]{revtex4-1}
\usepackage[utf8]{inputenc}
\usepackage{graphicx,color}
\usepackage{caption}
\usepackage{subcaption}
\usepackage{amsmath}
\usepackage{epstopdf}
\usepackage{soul}
\usepackage{tabularx}
\usepackage{romannum}
\newcommand{\D}{\mathrm{d}}

\newcommand{\half}{\frac{1}{2}}
\newcommand{\kro}{\delta_{\alpha\beta}}
\newcommand{\be}{\begin{equation}}
\newcommand{\ee}{\end{equation}}
\newcommand{\bea}{\begin{eqnarray}}
\newcommand{\eea}{\end{eqnarray}}
\newcommand{\ba} {\begin{align} }
\newcommand{\ea} {\end{align} }

\newcommand{\s}{\mathrm{s}}
\newcommand{\n}{\mathrm{n}}

\newcommand{\re}{\mathrm{Re}}

\newcommand{\vecr}{\boldsymbol{r}}
\newcommand{\vecf}{\boldsymbol{f}}
\newcommand{\vecq}{\boldsymbol{q}}
\newcommand{\vecp}{\boldsymbol{p}}
\newcommand{\vecv}{\boldsymbol{v}}
\newcommand{\vecj}{\boldsymbol{J}}
\newcommand{\vech}{\boldsymbol{h}}

\newcommand{\vecfrel}{\boldsymbol{f}^{\rm rel}}

\newcommand{\tenH}{\boldsymbol{H}^{B}}
\newcommand{\tenB}{\boldsymbol{B}}
\newcommand{\tenQ}{\boldsymbol{Q}}
\newcommand{\tenE}{\boldsymbol{E}}
\newcommand{\tenM}{\boldsymbol{M}}
\newcommand{\tensig}{\boldsymbol{\sigma}}
\newcommand{\delmu}{\Delta\mu}
\newcommand{\bmu}{\bar{\mu}}
\newcommand{\Ls}{L^{\rm s}_t}
\newcommand{\Ln}{L^{\rm n}_t}

\newcommand{\ra}[1]{\textcolor{black}{#1} } 

\begin{document}

\title{Permeation Instabilities in Active, Polar Gels}

\author{Ram M. Adar$^{1,2,3}$ and Jean-François Joanny$^{1,2,3}$}
\email{ram.adar@college-de-france.fr}
\affiliation{$^1$ Collège de France, 11 place Marcelin Berthelot, 75005 Paris, France\\
$^2$ Laboratoire Physico-Chimie Curie, Institut Curie, Centre de Recherche, Paris Sciences et Lettres Research University, Centre National de la Recherche Scientifique, 75005 Paris, France\\
$^3$ Université Pierre et Marie Curie, Sorbonne Universités, 75248 Paris, France
}


\begin{abstract}
We present a theory of active, permeating, polar gels, based on a two-fluid model. An active relative force between the gel
components creates a steady-state current. We analyze its stability, while considering  two polar coupling terms to the relative
current: a \ra{permeation-deformation} term, which describes network deformation by the solvent flow, and a permeation-alignment term, which describes the alignment of the polarization field by the network deformation and flow. Novel instability mechanisms emerge at finite
wave vectors, suggesting the formation of periodic domains and mesophases. Our results can be used to determine the physical conditions required for various types of multicellular migration across tissues.
\end{abstract}

\maketitle


{\it Introduction.} Active materials are driven out of equilibrium by a constant consumption of energy at the
microscopic level, which is converted into forces and motion~\cite{Marchetti13}. These include, among others,
biological objects on different scales, ranging from active motors, to living cells, and even groups
of animals. A useful framework for the study of active matter is hydrodynamics. Similarly to continuum theories of
liquid crystals~\cite{deGennesLC}, it describes macroscopic physical properties and flows, relying on
conservation laws and symmetries. It also provides an efficient language to distinguish between active materials,
based on their composition, orientational order, and rheological properties.

The biological motivation to our physical theory is multicellular migration. Connective tissues are made of cells in a complex extracellular environment, which often has a viscoelastic behavior~\cite{Levental07}. Cells may migrate collectively in tissues in a fluid-like manner~\cite{Hakim17}. We propose that the tissue can be regarded, therefore, as an {\it active, permeating, gel} with the cells acting as a solvent.
We further focus on a {\it polar} solvent, relevant to cells with spindle-like shapes and a preferred direction.
\ra{While active, permeating, polar gels have been studied in other contexts in the past~
\cite{CallanJones11,CallanJones13,Brand13,Pleiner16,Maitra19}, these studies remain at a general level, without interpreting the new Onsager transport coefficients of the theory, or clarifying the nature of the interaction between the two gel components.}

\ra{Our new theory is formulated in a systematic way as a two-fluid model. It identifies the internal forces of each component and the interaction forces between components}, which orient the solvent (“permeation alignment”) and deform the network (“permeation deformation”). These mechanisms drive novel, finite-wavelength instabilities, unique to active, permeating polar gels.  \ra{Our theory opens an avenue to study cell-matrix interactions during multicellular migration}.

{\it Theory.} We consider a two-component gel, composed of an active, polar solvent (s) and a viscoelastic
network (n). The polarization field is given by the unit vector $\vecp$. The network configuration is described by the left
Cauchy-Green strain tensor $\tenB=\tenE \tenE^{T}$, where $\tenE$ is the deformation gradient tensor. \ra{We consider the network component to be viscoelastic; elastic at short times and flowing at long times}. It has a volume fraction $\phi$ and the solvent $1-\phi$. The gel is assumed to be incompressible.

The free-energy of the gel can be decomposed into $F=\int\,\D\vecr\,\left(f_{p}+f_{B}+f_{Bp}+f_{\phi}\right)$, where
$f_p$ is the polarization free-energy density, $f_{B}$ is the elastic free-energy density, $f_{Bp}$ is a strain-polarization coupling term, and $f_{\phi}$ is the mixing free-energy density. The polarization contribution is given
by
\ba
\label{eq1}
f_{p}&=\left(1-\phi\right)^2\left[\half K\left(\nabla\vecp\right)^2+K_d\nabla\cdot\vecp\right]-\half h_\parallel\vecp^2.
\end{align}
It accounts for distortions of the polarization field around a fully polarized state~\cite{Kruse05,Voituriez06}. In
Eq.~(\ref{eq1}), $K$ is the Frank constant in the single-constant approximation and $K_d$ is a polar splay coefficient,
while $h_\parallel$ is a Lagrange multiplier to ensure that $\vecp^2=1$.
The polar splay term, $\left(1-\phi\right)^2 K_d \nabla\cdot\vecp$, is the only polar term in the free energy.
It plays an important role in our theory because of its coupling to the concentration; otherwise, it reduces to a
boundary term. The coupling is considered to scale as  $\left(1-\phi\right)^2$, because the free energy originates from
solvent-solvent interaction.

The gel is active. It is constantly driven out of equilibrium by the input of a fixed energy-density, $\Delta\mu$ that corresponds, for example, to the chemical-potential difference between ATP and its hydrolysis products~\cite{Prost15,Joanny07}.

We describe the dynamics of the concentration, polarization, and strain within a hydrodynamic
framework. The network moves with a velocity $\vecv^\n$ and the solvent with a velocity $\vecv^\s$, corresponding to a center-of-mass (COM) velocity, $\vecv=\phi\,\vecv^\n+\left(1-\phi\right)\vecv^\s$, and a relative current, $\vecj=\phi\left(1-\phi\right)
\left(\vecv^\n-\vecv^\s\right)$. We have assumed, for simplicity, the same molecular mass for both components.

The dynamics of the concentration are determined from the continuity equation, $\partial_t \phi+\nabla\cdot\left(\phi\,\vecv^\n\right)=0.$ For the polarization and network configuration, we derive in the Supplemental Material (SM)~\cite{SI} the following, minimal constitutive relations:
\ba
\left(\partial_t+\vecv^\s\cdot\nabla\right)\vecp &=\frac{1}{\gamma_1} \vech+\vecp\cdot\nabla\vecv^\s+\lambda\vecj,\label{eq2}\\
\left(\partial_t+\vecv^\n\cdot\nabla\right)\tenB &=-\frac{1}{\tau} \frac{\partial \tenB}{\partial \tensig^{\rm el}}:\tensig^{\rm el}+\tenB\nabla\vecv^\n+\left(\nabla\vecv^\n\right)^T\tenB\nonumber\\
&+\half\xi\left(\vecj\vecp+\vecp\vecj\right).\label{eq3}
\end{align}
In Eq.~(\ref{eq2}), $\gamma_1$ is the rotational viscosity, $\vech=-\delta F/\delta\vecp$ is the solvent
orientational field, and the  second term in the right-hand side (RHS) is a convective term~\cite{Lie,holzapfel,Hemingway14,Hemingway16}. In Eq.~(\ref{eq3}), 
 $\tau$ is a viscoelastic relaxation time and $\tensig^{\rm el}$ is the elastic (Kirchhoff) stress~\cite{holzapfel}. It is given by $\tensig^{\rm el}=-2\tenH\tenB$, where
$\tenH=-\delta F/\delta\tenB$ is the network molecular field.
The next two terms in Eq.~(\ref{eq3}) are convective terms~\cite{Lie}.

\begin{figure}[ht]
\centering
\includegraphics[width=0.99\columnwidth]{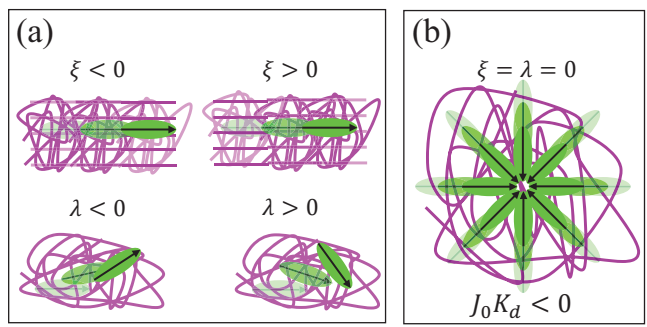}
\caption{(Color online) Heuristic description of a polar solvent (green, polarization indicated by a black arrow) and a viscoelastic network (purple). (a) Reactive, polar couplings for $J_0<0$; \ra{permeation-deformation} coupling, where the network becomes more aligned ($\xi<0$) or less aligned ($\xi>0$) with the relative current and network polarization, and permeation-alignment coupling, where the solvent becomes aligned against ($\lambda<0$) or in the direction of the relative current ($\lambda>0$). (b)
The system is unstable for $\xi=\lambda=0$ and $J_0 K_d<0$, where the relative current brings the polar solvent closer together and increases its concentration. 
}
\label{fig1}
\end{figure}

The last terms in RHS of Eqs.~(\ref{eq2}) and (\ref{eq3}) are reactive
couplings allowed by the polar symmetry. We refer to $\lambda$ as the {\it permeation-alignment} parameter. It couples the polarization rate with the relative current. We refer to $\xi
$ as the {\it \ra{permeation-deformation}} parameter. It couples the network strain-rate with the relative
current. Both $\lambda$ and $\xi$ have units of inverse length. They are central to our
work, and we give a heuristic description of their roles in Fig.~\ref{fig1}a.  In the absence of
polarization and for a linear elastic stress-strain relation, Eq.~(\ref{eq3}) reduces to the upper-convected Maxwell equation~\cite{Larson}.

Onsager's reciprocal relations infer reciprocal, reactive couplings involving $\lambda$ and $\xi$ in
the constitutive equation for the relative current, $\vecj$. As in the two-fluid model, friction due to the
relative current acts as a relative force between the components. Therefore, the new
permeation couplings are concurrent with new relative forces between the gel components~\cite{SI},
\be
\label{eq4}
\vecfrel=\frac{1}{\gamma}\vecj-\phi\left(1-\phi\right)\left(\lambda\vech+\xi\tenH\cdot\vecp+\nu\Delta\mu\vecp\right).
\ee
Here we included an active relative force $\sim\nu\Delta\mu$, where $\nu$ has units of inverse length, which results in an active relative current.

Overall, the force-balance equations for the two components read
\ba
\label{eq5}
\vecf^{\n}-\phi\nabla\delta P &=\vecfrel,\nonumber\\
\vecf^{\s}-\left(1-\phi\right)\nabla\delta P&=-\vecfrel,
\end{align}
where $\vecf^{\n}$ and $\vecf^{\s}$ are the forces acting on the network and solvent, respectively, and $\delta P$ is a pressure difference that enforces incompressibility~\cite{SI}. Equation~(\ref{eq5})
reduces to a standard two-fluid model~\cite{Onuki92} in the absence of activity and polarization. As
the new relative forces do not include any derivatives, as opposed to the stress and pressure terms, they are especially important
in the limit of small wave vectors.

The forces acting on each of the components are~\cite{SI}
\ba
f^{\n}_\alpha&=\partial_\beta\sigma^{\rm el}_{\alpha\beta}-\phi\partial_\alpha\bmu-H^B_{\beta\gamma}\partial_\alpha B_{\gamma\beta},\label{eq6} \\
f^{\s}_\alpha&=\partial_\beta\left[2\eta_s v^s _{\alpha\beta}-h_\alpha p_\beta+\left(1-\phi\right)\zeta\Delta\mu Q_{\alpha\beta}\right]-h_\beta\partial_\alpha p_\beta\label{eq7}.
\end{align}
In Eq.~(\ref{eq6}), 
the second term in RHS is the osmotic pressure gradient with $\bmu=\delta F/\delta\phi$ being the relative chemical potential, and the last term originates in the Ericksen stress of the gel. In Eq.~(\ref{eq7}), $\eta_s$ is the solvent viscosity and $v^\s _{\alpha\beta}=\left(\partial_\alpha v^\s _\beta+\partial_\beta v^\s _\alpha\right)/2$ is the solvent strain rate. The next term is the stress due to polarization rotations and the last term in the parenthesis is an active stress, proportional to the nematic tensor, $\tenQ$, and solvent concentration. 
The last term in RHS also originates in the Ericksen stress. These equations satisfy Onsager reciprocity with the convective terms in Eqs.~(\ref{eq2}) and (\ref{eq3}).

{\it Linear stability analysis.} 
We examine the linear stability of the steady state with respect to perturbations with a growth rate $s$ and wave vector $q$, of the form $\exp\left(st+i\vecq\cdot\vecr\right)$. The steady state is homogeneous, 
$\phi=\phi_0$, $\vecp=\vecp^0=\hat{x}$, and $\tenB=\tenB^0$, with a relative current driven by the active relative force, given by $\vecj^0=J_0\,\vecp^0$ with $J_0=\gamma \phi_0\left(1-\phi_0\right)\nu\delmu$. The system is stable if $\re\, s<0$ for all the eigenvalues of the linear system. The details of the analysis are found in the SM~\cite{SI}.
For simplicity and in order to highlight new instabilities that result from the polar
couplings, we focus on a 2-dimensional system 
with wave vectors perpendicular to the steady-state polarization, $
\vecq=q\hat{y}$~\cite{shearinstability,Mishra06,voituriez05}. We consider the strain
free-energy, $f_B=G\phi{\rm Tr}\left(\tenB-\ln \tenB\right)/2$, corresponding to Gaussian polymer
chains~\cite{Milner93,Flory,strain}, where $G\phi$ is the shear modulus, and $f_{Bp}=0$. 

In the hydrodynamic limit, we consider small wave vectors and solve for the growth
rate up to quadratic order in $q$, $s\simeq s_0+iuq-
Dq^2$, where $s_0$ is a relaxation rate, $u$ a velocity, and $D$ a diffusion coefficient.  In the opposite, large-$q$ limit, the system is always stable~\cite{SI}.

\begin{figure*}[ht]
\includegraphics[width=0.295\textwidth]{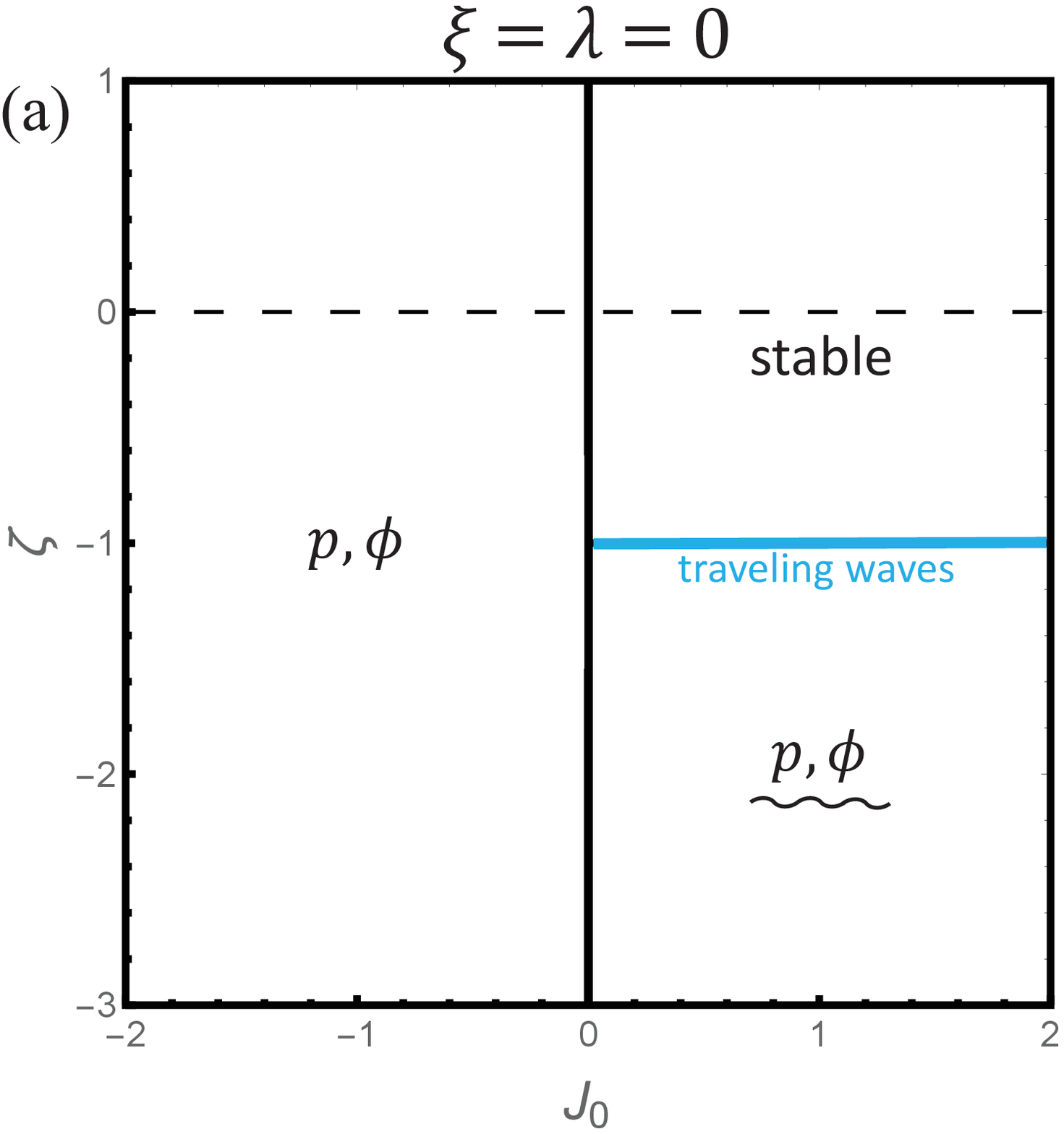} \hspace{0.5cm}
\includegraphics[width=0.3\textwidth]{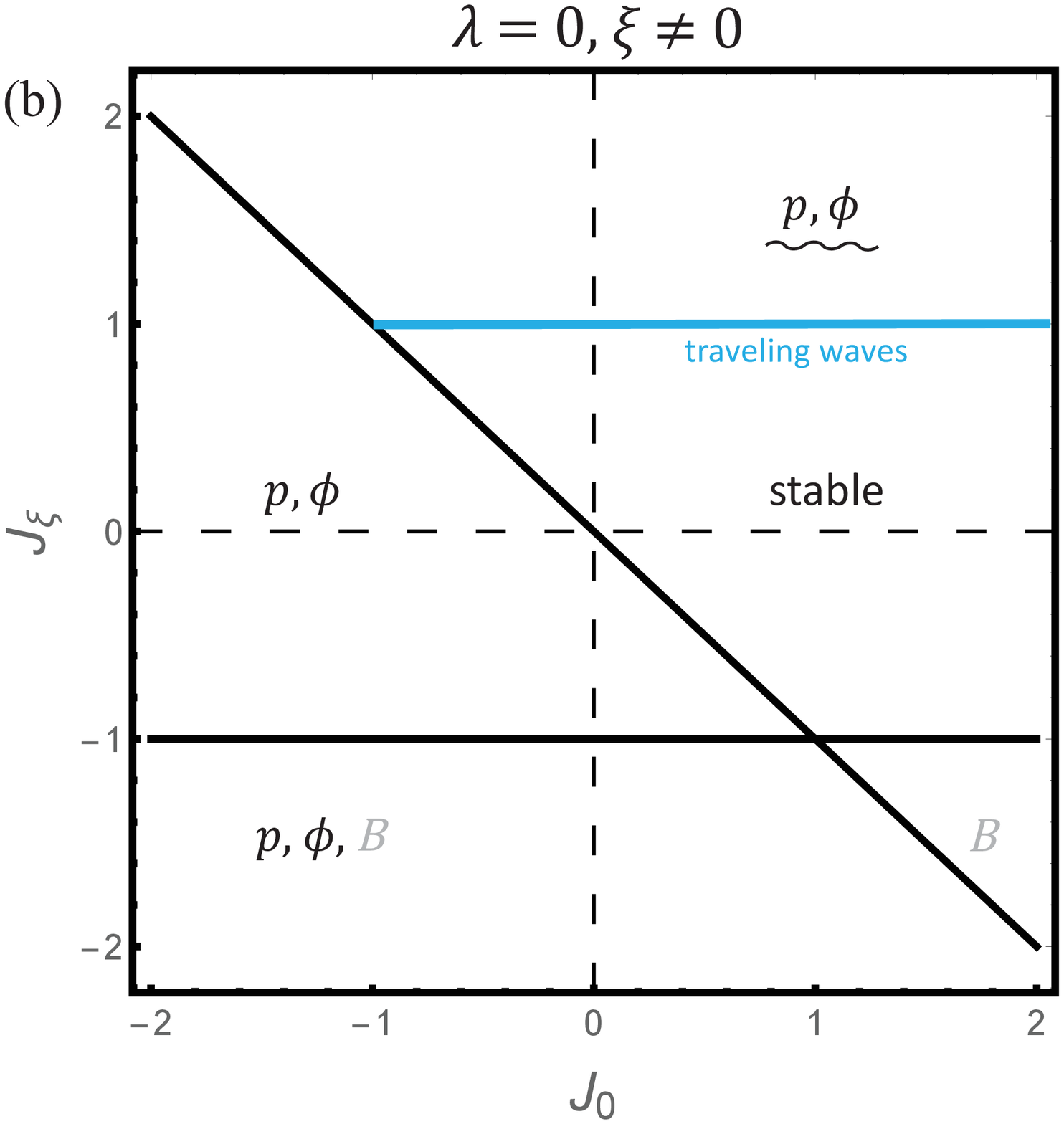}\hspace{0.5cm}
\includegraphics[width=0.3\textwidth]{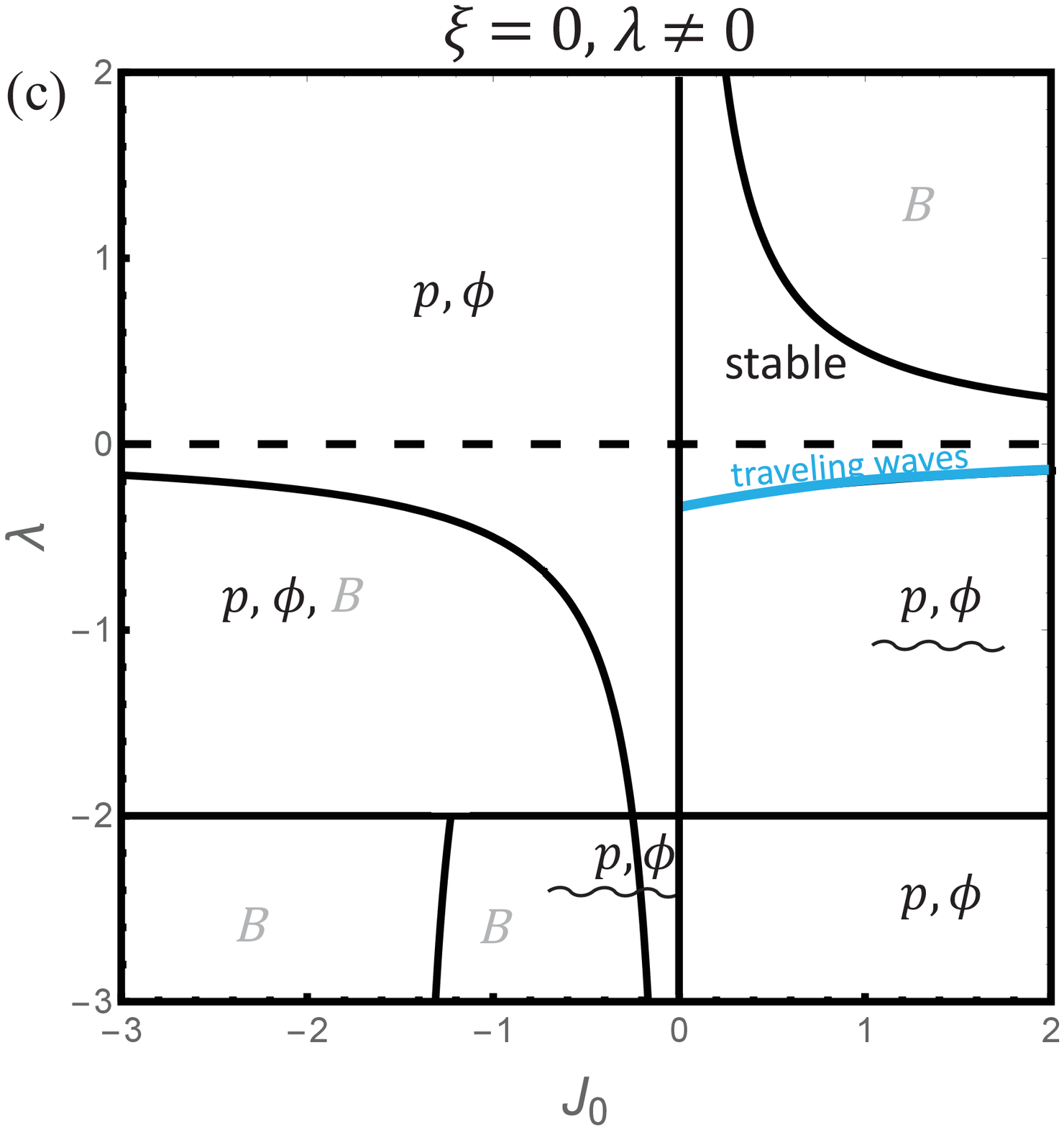}
\caption{(Color online) \ra{Linear} stability diagrams for active, permeating, polar gels. Polarization, concentration, and possible strain instabilities are denoted by $p,\,\phi$ and grey $B$, respectively. Waves indicate instabilities that oscillate in time. $J_0$ is written in units of $\left(1-\phi_0\right)l_p/\tau_p.$ (a) Stability diagram for $\xi=\lambda=0$. The value $2\left(D_p+D_\phi\right)=\gamma\phi_0\Delta\mu $ is used. (b) Stability diagram for $\lambda=0,\,\xi\ne0$ with $J_\xi$ written in units of $\left(1-\phi_0\right)l_p/\tau_p$. The values $1+\left(D_\phi+D_\zeta\right)/D_p=D_B/D_p=2\eta_s\tau/\left(\eta_s+\eta_n\right)\tau_p$ are used. (c) Stability diagram for $\xi=0,\,\lambda\ne0$ with $\lambda$ written in units of $1/\left(1-\phi_0\right)l_p$. The values $D_p=D_\phi+D_\zeta$, $4\gamma_1/\left(1-\phi_0\right)=3\left[\eta_s/\left(1-\phi_0\right)^2+\eta_n/
\phi_0^2\right]$,  $2\phi_0\gamma_1\gamma=3\left(1-\phi_0)\right)l_p^2$, and $\tau=2\tau_p$ are used.}
\label{fig2}
\end{figure*}

First, we analyze the stability for $\lambda=\xi=0$.  The uniform steady-state has no deformation (isotropic network),
$B^0_{\alpha\beta}=\delta_{\alpha\beta}$. There
are two purely hydrodynamic modes with $s_0=0$, which correspond to linear combinations of $p^1$ and $\phi^1$ for $q=0$. Their velocity is given by $u=\pm\sqrt{2J_0l_p/\left[\tau_p\left(1-
\phi_0\right)\right]}$, where $l_p=K/K_d$ is a polarization length-scale, which may be negative,
and $\tau_p=l_p^2/D_p$  is a relaxation time associated with the angular diffusion coefficient,
$D_p=\left(1-\phi_0\right)^2 K/\gamma_1$. The velocity is imaginary for $J_0K_d<0$, in which
case the growth rate is positive. This instability can be understood intuitively; the polar splay
aligns the solvent molecules towards each other, while the active relative current brings them
closer together, as is illustrated in Fig.~\ref{fig1}b.

The quadratic correction is given by $D=\left(D_p+D_\phi+D_\zeta
\right)/2$, where $D_\phi=\gamma\phi_0\left(1-\phi_0\right)/
\chi$ is the osmotic diffusion coefficient, with $1/\chi=\partial^2f_{\phi}/\partial\phi^2$ being the inverse osmotic compressibility. The active term, $D_\zeta=\phi_0\gamma\zeta\Delta\mu/2$,  originates in the concentration dependence of the active stress. The active stress varies with the concentration, resulting in a relative current that modifies the concentration further. 
For sufficiently negative active stresses, the quadratic correction vanishes and then becomes negative. The critical active stress when this occurs is $\zeta_{\rm c}\Delta\mu=-2\left(D_p+D_\phi\right)/\left(\gamma \phi_0\right)$.

The system is unstable for a combination of an imaginary $u$ and negative $D$, where the growth rate is positive and increases with $q$. As the system is stable for large wavenumbers, this instability persists only
up to a finite $q$, and there exists a most unstable wave vector, $q^\ast$,  with a fastest growth rate, $s^\ast$. For an imaginary velocity, $u=i|u|$, and positive quadratic coefficient, $D>0$, they are found analytically as $q^\ast=|u|/2D$ and $s^\ast=|u|^2/4D$. If the velocity $u$ is real, a vanishing diffusion constant ($D=0$) infers traveling waves
(Hopf bifurcation). Beyond this threshold, for $D<0$, the concentration-polarization instability
is oscillating in time, and the values of $q^\ast$ and $s^\ast$ can be calculated numerically. The \ra{linear stability analysis} for $\xi=\lambda=0$ is summarized in Fig.~\ref{fig2}a. Note that in the passive limit ($\delmu=0$), the linear term vanishes ($u=0$), and the system is unstable for $4\chi K_d^2>K$~\cite{Voituriez06}. We assume that $4\chi
K_d^2<K$ hereafter.

Next, we perform the linear stability analysis for $\xi\ne0$ and $\lambda\ne0$. In addition to a polarization-concentration instability,
we demonstrate a possible strain instability. The eigenvector of this instability reduces to a strain component for $q=0$ ($B_{xy}$ for $\xi\ne0$ and $B_{yy}$ for $\lambda\ne0$), and its growth rate is $s=s_0-D q^2$ with $s_0<0$. As the growth rate is negative for both small and large
$q$ values in this case, a numerical calculation of $s(q)$ is required to verify the instability for intermediate $q$ values.

{\bf \ra{Permeation deformation} $\left(\lambda=0,\,\xi\ne0\right)$}. The \ra{permeation-deformation} coupling, combined with the active relative current, deforms the network in the steady state, $B^0_{\alpha\beta}=\kro+\xi J_0\tau p^0_\alpha p^0_\beta$. The network is
more (less) aligned with the flowing solvent for $\xi J_0>0$ ($\xi J_0<0$). The network also expands (contracts) for  $\xi J_0>0$ ($\xi J_0<0$). As $B_{\alpha\beta}$ is a positive-definite tensor, a steady state exists only for $\xi J_0 \tau >-1$. We assume a small value of $\xi J_0 \tau$ and expand our results to linear order in $\xi$~\cite{SI}.

The \ra{permeation-deformation} coupling retains the  possible polarization-concentration instability to linear order in $q$, with
$u=\pm\sqrt{2\left(J_0+J_\xi\right)l_p/\left[\tau_p\left(1-\phi_0\right)\right]}$. Compared to our
previous result, note the additional active relative-current term, $J_\xi=\xi\gamma\phi_0\left(1-
\phi_0\right)^{2}\zeta\delmu\eta_n/\left[2\left(\eta_s+\eta_n\right)\right]$, where $\eta_n=G\phi_0\tau
$ is the network viscosity. This current originates in the active stress, which strains the
network, and induces a relative current due to the \ra{permeation-deformation} coupling. An instability
occurs for $\left(J_0+J_\xi\right)K_d<0$.

The diffusion coefficient is given by $D=\left(D_p+D_\phi +D_\zeta+D_\xi\right)/2$, with $D_\xi=-2\eta_s
\tau l_p J_\xi/\left[\left(1-\phi_0\right)\left(\eta_s+\eta_n\right)\tau_p\right]$, it can be either
positive or negative, depending on the sign of $J_\xi K_d$. The mechanism driving the
instability can be understood by considering a small concentration fluctuation. The polar-splay term results in a polarization rotation that strains the network, due to the active stress.
The \ra{permeation-deformation} coupling then induces a relative current that modifies the
concentration. The feedback can be either positive or negative.

The \ra{permeation-deformation} coupling may lead to a shear-strain instability as well.
The shear strain relaxes at $q=0$ with a rate $s_0=-\left(1+\eta_n/\eta_s\right)/\tau$. The linear
correction vanishes, while the diffusion coefficient is given by $D=D_B-D_\xi$, where $D_B=G
\gamma/\left(1-\phi_0\right)$ is the strain diffusion coefficient, due to permeation. This infers a
possible instability for $D_\xi>D_B$. The mechanism driving the instability is as follows: a shear strain induces a relative current, due to \ra{permeation deformation}. The resulting concentration gradient rotates the polarization due to the polar splay
term, and the resulting active stress shears the network further. This feedback can be either
positive or negative.

The \ra{linear stability analysis} in the presence of \ra{permeation deformation} is summarized
in Fig.~\ref{fig2}b. As the instabilities are mainly related to network deformations, stability can be achieved
by reducing strain. This is possible either by a strong suppression  (large $G$ with fixed $\tau
$) or fast relaxation (small $\tau$ for fixed $G$).

{\bf Permeation alignment $\left(\xi=0,\,\lambda\ne0\right)$}. We study the stability up to linear order in $\lambda$~\cite{SI}. The network is isotropic in the steady-state with $B^0_{\alpha\beta}=\delta_{\alpha\beta}$. The polarization-concentration velocity is $u=
\pm\sqrt{\left[2l_p/\tau_p\left(1-\phi_0\right)+\lambda \left(D_\phi+D_\zeta\right)\right]J_0}$. The second term in the parenthesis
is independent of polar splay; as the polarization rotates, it exerts an active relative force, which
leads to a relative current. The permeation-alignment coupling then rotates the polarization
further. An instability to linear order in $q$  occurs when the argument of the square root is
negative.

The diffusion coefficient is given by  $D=\left(D_p+D_\phi+D_\zeta+D_\lambda\right)/2$, with $D_
\lambda=2\lambda D_\phi\chi\left[\left(\eta_s/\left(1-\phi_0\right)^2+\eta_n/
\phi_0^2\right)J_0+2\phi_0\left(1-\phi_0\right)^2 K_d\right]$. It can be either positive or
negative and includes contributions from two mechanisms:  any polarization fluctuation
causes both an active relative force $\sim J_0$ (first mechanism) and a concentration gradient
through the polar-splay coupling $\sim K_d$ (second mechanism). Both induce a relative current
that rotates the polarization, due to the permeation alignment coupling. This feedback can be
either positive or negative.

The permeation-alignment coupling can result in an instability for the elongation strain, $B_{yy}
$. The growth rate relaxes for $q=0$ as $s_0=-1/\tau$. The linear correction vanishes, while the
diffusion coefficient is given by $D=\left(1-\lambda\tau J_0\right)D_B$. In order to understand
the $\lambda$ term, consider a fluctuation in $B_{yy}$.
The resulting stress induces a relative current that rotates the polarization by permeation alignment. The active relative force then modifies the relative current that further strains the
network by convection.

The \ra{linear stability analysis} in the presence of permeation alignment is summarized in
Fig.~\ref{fig2}c. As the instabilities are mainly related to the relative current in the $y$-direction, stability
can be achieved by lowering the pressure gradient. This is possible by lowering the solvent and
network viscosities, which induce smaller pressures.


{\it Discussion.}
%
In this Letter, we have reported finite-wavelength instabilities that result from polar couplings to the relative current between a viscoelastic network and active, polar fluid. This implies possible mesophases and periodic domains with continuous flow patterns ~\cite{Voituriez06,Blankschtein85,Hinshaw88}.  The permeation couplings may also modify known instabilities in ordered, active nematics~\cite{voituriez05,Hemingway14,Hemingway16},  close to the isotropic-polar transition~\cite{Markovich}, \ra{and in the shape of active permeating sheets~\cite{Ideses2018}}.  

Our theory can be used to describe cell migration in tissue. Cells often migrate collectively in a
fluid-like manner with weak and short-lived mutual adhesions (``multicellular streaming''~
\cite{Hakim17,Friedl09}). In a coarse-grained view, this can be regarded as permeation of an
active, polar fluid in a viscoelastic network. Our analysis suggests the required
physical conditions for migrating cells to traverse a tissue homogeneously (a stable, flowing steady-state),
as opposed to  \ra{migration in strands or} local cell movement in confined domains (finite-$q$ instabilities).

The novel ingredients of our theory describe in this context the forces exerted between cells and, for example, the extra-cellular matrix (ECM), including matrix deformation \ra{~\cite{Sahai20,Danijela21}}. The cross-talk between migrating cells and the ECM is called ``dynamic reciprocity'', and it is considered important to embryonic development, tissue regeneration, and metastasis~\cite{Alexander16,vanHelvert18,Clark15}. \ra{Our work thus provides a new, hydrodynamic framework to describe ``dynamic reciprocity'' during collective migration}. We investigate this application further in a separate study~\cite{nextpaper}.

{\it Acknowledgements.} R. M. A. acknowledges support from Yad Hanadiv through a Rothschild Fellowship and from ANR Grant No. ANR-18-CE30-0005. We thank L. Truskinovsky, D. Grossman, \ra{Matthieu Piel, Danijela Vignjevic, and Erik Sahai} for fruitful discussions.

 \small


\onecolumngrid
\newpage
\appendix
\section*{Permeation Instabilities in Active, Polar Gels: Supplemental Material}
This Supplemental Material (SM) provides, in greater detail, the derivation
of the dynamic equations and calculations that are relevant to the
linear stability analysis. The outline of the SM is as follows. In
Sec.~\Romannum{1}, Equations (2)-(7) of the Letter are derived, using the general framework of non-equilibrium thermodynamics. In Sec.~~\Romannum{2}, the equations are written explicitly in
terms of the concentration, polarization, and strain variables, and
are linearized around the steady state. Next, in Sec.~\Romannum{3}, we review
the parameters that are used in our theory and relate them to convenient
lengthscales and timescales. Then, we focus on the linear stability
analysis. The stability of the system in the large-$q$ limit is demonstrated
in Sec.~\Romannum{4}. As part of Sec.~\Romannum{5}, we detail the approximations used throughout
our work. Finally, in Sec.~\Romannum{6}, we analyze the linear stability
in the passive case.

\section{\Romannum{1}. Derivation of the dynamic equations and force balance equations}
\label{SMsec1}
In this Section we derive the dynamic equations [Eqs.~(2)-(3) in the main text] and force-balance equation [Eqs.~(4)-(7) in the main text] from the general framework of non-equilibrium thermodynamics. First, the free energy production rate is written in a convenient form, and constitutive relations are written, while respecting Onsager reciprocity. Next, we demonstrate how these equations can yield a two-fluid model.
\subsection{A. Derivation of the dynamic equations}
The time derivative of the free energy is given by (see similar cases in~\cite{CallanJones11,Pleiner16})
\begin{align}
\label{eqSprod1}
\dot{F} & =-\int \D\boldsymbol{r}\left[v_{\alpha\beta}\sigma_{\alpha\beta}^{\D}+h_{\alpha}\frac{D}{Dt}p_{\alpha}+H^B_{\alpha\beta}\frac{D}{Dt}B_{\alpha\beta}-J_{\alpha}\partial_{\alpha}\bmu+\Delta\mu r\right],
\end{align}
where $v_{\alpha\beta}=\left(\partial_{\alpha}v_{\beta}+\partial_{\beta}v_{\alpha}\right)/2$
is the center-of-mass (COM) strain rate and $\sigma_{\alpha\beta}^{\D}$ is the symmetric, deviatoric
stress tensor. The solvent orientational field is $h_{\alpha}=-\delta F/\delta p_{\alpha},$
while the co-rotational derivative of the polarization is given by
$Dp_{\alpha}/Dt=\left(\partial_{t}+v_{\beta}\partial_{\beta}\right)p_{\alpha}+\omega_{\alpha\beta}p_{\beta}$,
with $\omega_{\alpha\beta}=\left(\partial_{\alpha}v_{\beta}-\partial_{\beta}v_{\alpha}\right)/2$
being the COM vorticity tensor. The network molecular field is $H^B_{\alpha\beta}=-\delta F/\delta B_{\alpha\beta},$
while the co-rotational derivative of the strain tensor is $DB_{\alpha\beta}/Dt=\left(\partial_{t}+v_{\gamma}\partial_{\gamma}\right)B_{\alpha\beta}+\omega_{\alpha\gamma}B_{\gamma\beta}+\omega_{\beta\gamma}B_{\gamma\alpha}.$
As $B_{\alpha\beta}$ is a symmetric tensor, so is $H^B_{\alpha\beta}.$
The relative current between the two components is $J_{\alpha}=\phi\left(1-\phi\right)\left(v_{\alpha}^{n}-v_{\alpha}^{s}\right),$ while the relative chemical potential is $\bmu=\delta F/\delta\phi.$ Finally, $r$
is the rate associated with the active consumption of the energy density
$\Delta\mu$.

The deviatoric stress tensor is related to the total stress tensor, $\sigma_{\alpha\beta}$, by
\begin{align}
\sigma^{\D}_{\alpha\beta}=\sigma_{\alpha\beta}-\sigma^a_{\alpha\beta}+\rho \,v_\alpha v_\beta-\sigma^{\rm Er, s}_{\alpha\beta},
\end{align}
where $\sigma^a_{\alpha\beta}$ is the antisymmetric part of the total stress and $\rho v_\alpha v_\beta$ is the momentum transfer with $\rho$ being the total mass density. We neglect this term hereafter.  The last term in the equation above is the symmetric part of the Ericksen stress. The Ericksen stress tensor, $\sigma^{\rm Er}_{\alpha\beta}$, is given by~\cite{Joanny07}
\be
\sigma^{\rm Er}_{\alpha\beta}=\left(f-n_{\s}\mu_{\s}-n_{\n}\mu_{\n}\right)\delta_{\alpha\beta}-\frac{\partial f}{\partial\left(\partial_\beta p_\gamma\right)}\partial_\alpha p_\gamma-\frac{\partial f}{\partial\left(\partial_\beta n_\s\right)}\partial_\alpha n_\s-\frac{\partial f}{\partial\left(\partial_\beta n_\n\right)}\partial_\alpha n_\n,
\ee
where $n_{\rm n}$ and $n_{\rm s}$ are the network and solvent densities, respectively, and $\mu_{\rm n}$ and $\mu_{\rm s}$ are their chemical potentials.

It is customary to consider that the solvent and network components are each convected with their
own velocity~\cite{Milner93}. We, therefore, rewrite the free-energy production rate of Eq.~(\ref{eqSprod1})
in terms of a solvent convected derivative, $\Ls$,
and a network convected derivative, $\Ln$, defined
as\ra{
\begin{align}
\label{SMeqLt}
L_{t}^{\s}\,p_{\alpha} & =\left(\partial_{t}+v_{\beta}^{\s}\partial_{\beta}\right)p_{\alpha}+\omega^\s_{\alpha\beta}p_\beta+\nu v^\s_{\alpha\beta}p_\beta,\\
L_{t}^{\n}B_{\alpha\beta} & =\left(\partial_{t}+v_{\gamma}^{\n}\partial_{\gamma}\right)B_{\alpha\beta}+\omega^\n_{\alpha\gamma}B_{\gamma\beta}+\omega^\n_{\beta\gamma}B_{\gamma\alpha}+\nu_1\left(v^\n_{\alpha\gamma}B_{\gamma\beta}+v^\n_{\beta\gamma}B_{\gamma\alpha}\right)+A_{\alpha\beta},
\end{align}
where $\omega^\s_{\alpha\beta}$ and $\omega^\n_{\alpha\beta}$ are the solvent- and network vorticity tensors, respectively, while $v^\s_{\alpha\beta}$ and $v^\n_{\alpha\beta}$ are the solvent- and network strain-rate tensors, respectively. The term $\nu$ is the solvent shear-alignment parameter and $\nu_1$ is (minus) the network's slip parameter. The tensor $A_{\alpha\beta}$ accounts for other geometric non-linearities, according to
\begin{align}
\label{AeqSM}
A_{\alpha\beta}=\nu_2 v^\n_{\gamma \gamma}B_{\alpha\beta}+\nu_3 v^\n_{\gamma\gamma}B_{\delta\delta}\delta_{\alpha\beta}+\nu_4 B_{\gamma\gamma}v^\n_{\alpha\beta}+\nu_5 v^\n_{\gamma\delta}B_{\gamma\delta}\delta_{\alpha\beta}.
\end{align}
Note that other terms are allowed by the polar symmetry, such as those containing contractions between $Q_{\alpha\beta}$ and either $B_{\alpha\beta}$ or $v^\n_{\alpha\beta}$. For simplicity, we set $\nu=\nu_{1}=-1$ and $A_{\alpha\beta}=0$, such that
\begin{align}
\label{SMeqLt}
L_{t}^{\s}\,p_{\alpha} & =\left(\partial_{t}+v_{\beta}^{\s}\partial_{\beta}\right)p_{\alpha}-p_{\beta}\partial_{\beta}v_{\alpha}^{\s},\\
L_{t}^{\n}B_{\alpha\beta} & =\left(\partial_{t}+v_{\gamma}^{\n}\partial_{\gamma}\right)B_{\alpha\beta}-B_{\gamma\beta}\partial_{\gamma}v_{\alpha}^{\n}-B_{\gamma\alpha}\partial_{\gamma}v_{\beta}^{\n}.
\end{align}
 The convected derivatives $\Ls$ and $\Ln$ thus reduce to vector and tensor Lie derivatives, respectively~\cite{holzapfel}. In particular, $\Ln B_{\alpha\beta}$ is the upper-convected derivative~\cite{Larson}. For $\nu_1=1$ and $A_{\alpha\beta}=0,$ the network convected derivative reduces to the lower-convected derivative.}

Inserting the convected derivatives and using integration by parts, Eq.~(\ref{eqSprod1}) transforms into
\begin{align}
\label{eqSprod2c}
\dot{F}&=-\int \D\boldsymbol{r}\left[v_{\alpha\beta}\delta\sigma_{\alpha\beta}+h_{\alpha}\Ls\, p_{\alpha}+H^B_{\alpha\beta}\Ln B_{\alpha\beta}+\Delta\mu r \right.\nonumber\\ &\left.+J_{\alpha}\left(-\partial_{\alpha}\bmu+\frac{1}{1-\phi}\left(h_\beta \partial_\alpha p_\beta+\partial_\beta\left(h_\alpha p_\beta\right)\right)+\frac{1}{\phi}\left(\partial_\beta\sigma^{\rm el}_{\alpha\beta}-H^B_{\beta\gamma}\partial_\alpha B_{\beta\gamma}\right)\right)\right].
\end{align}
In Eq.~(\ref{eqSprod2c}), we have defined the stress
\begin{align}
\label{eqSMdelsig}
\delta\sigma_{\alpha\beta}= \sigma_{\alpha\beta}^{\D}-\half\left(\sigma^{\rm el}_{\alpha\beta}+\sigma^{\rm el}_{\beta\alpha}-h_\alpha p_\beta-h_\beta p_\alpha\right),
\end{align}
where $\sigma^{\rm el}_{\alpha\beta}=-2H^B_{\alpha\gamma}B_{\gamma\beta}$ is the elastic (Kirchhoff) stress~\cite{holzapfel} and $-h_\alpha p_\beta$ is the stress associated with polarization rotations. These are the reactive contributions that result from the convected derivatives.

The stress $\delta\sigma_{\alpha\beta}$ includes additional contributions to the deviatoric stress and, namely, the dissipative and active stress. As the network is viscoelastic, dissipation due to network viscosity is already included in the $H^B_{\alpha\beta}\Ln B_{\alpha\beta}$ term, 
and the viscous contribution to $\delta\sigma_{\alpha\beta}$ originates from the solvent alone. For this reason, it is convenient to rewrite the free-energy production rate in terms of the solvent strain rate, $v^\s_{\alpha\beta}=\left(\partial_{\alpha}v^\s_{\beta}+\partial_{\beta}v^\s_{\alpha}\right)/2$, as
\begin{align}
\label{eqSprod2a}
\dot{F}&=-\int \D\boldsymbol{r}\left[v^\s_{\alpha\beta}\delta\sigma_{\alpha\beta}+h_{\alpha}\Ls\, p_{\alpha}+H^B_{\alpha\beta}\Ln B_{\alpha\beta}+J_\alpha\delta f_\alpha+\Delta\mu r \right].
\end{align}
The conjugate of the relative current, $\delta f_\alpha$, is a force density, given by
\begin{align}
\label{eqSMnewfluxes}
\delta f_\alpha&=\frac{1}{\phi}\left(\partial_\beta\sigma^{\rm el}_{\alpha\beta}-\phi\partial_{\alpha}\bmu-H^B_{\beta\gamma}\partial_\alpha B_{\beta\gamma}\right)-\frac{1}{1-\phi}\left[\partial_\beta \left(\delta\sigma_{\alpha\beta}-h_\alpha p_\beta\right)-h_\beta \partial_\alpha p_\beta\right].
\end{align}
We interpret it below in Sec~\ref{SMssec12}.

The free-energy production rate of Eq.~(\ref{eqSprod2a}) is written as an integral over pairs of
forces and conjugate fluxes. In each pair, we consider the first variable
as the force, and the second as the flux. In particular, the choice
of $J_{\alpha}$ as the force and of $\delta f_{\alpha}$
as the flux is in contrast to their physical units. This choice is
more convenient, because $J_{\alpha}$ is more easily measurable and
it appears in convected derivatives.

Our aim is to derive constitutive relations between forces and fluxes
in a linear theory, close to equilibrium. Fluxes are related to forces
with the same signature under time-reversal as their conjugate force,
by {\it dissipative} couplings, and to forces with
opposite signatures by {\it reactive} couplings.
Reciprocal dissipative couplings are equal and have a positive contribution
to the entropy production, while reciprocal reactive couplings have
opposite signs and do not contribute to the entropy production~\cite{deGroot}.

We consider for the constitutive relations only the leading, zeroth-order terms in a gradient expansion. There are fifteen such coupling terms (number of independent terms in a symmetric $5\times5$ matrix). One of which, relating $r$ and $\Delta\mu$, does not play any role in the dynamics of the gel. Below, we address the remaining terms and explain how we retain only eight of them. For simplicity, we consider scalar couplings.  More complicated tensors, written in terms of $p_{\alpha}p_{\beta}$ and $B_{\alpha\beta}$ are generally
applicable.

The polarization rate and strain rate are given by
\begin{align}
\Ls\,p_{\alpha}^{\left(d\right)} & =\frac{1}{\gamma_{1}}h_{\alpha},\nonumber\\
\Ls\,p_{\alpha}^{\left(r\right)} & =\lambda J_{\alpha}, \label{eqp}\\
\Ln B_{\alpha\beta}^{\left(d\right)} & =-\frac{1}{\tau}\frac{\partial B_{\alpha\beta}}{\partial\sigma^{\rm el}_{\gamma\eta}}\sigma^{\rm el}_{\gamma\eta},\nonumber\\
\Ln B_{\alpha\beta}^{\left(r\right)} & =\frac{1}{2}\xi\left(J_{\alpha}p_{\beta}+J_{\beta}p_{\alpha}\right),\label{eqB}
\end{align}
where $\left(d\right)$ denotes the dissipative part of the flux and
$\left(r\right)$ its reactive part.  The dissipative, diagonal term in the polarization rate is written in terms of the angular viscosity, $\gamma_1$. We have chosen the dissipative couplings to $\Delta\mu$ and $H^B_{\alpha\beta}$ to be zero. Both these terms are, to linear order, of the form $\sim B_{\alpha\beta}p_{\beta}$ (we disregard a term
$\Delta\mu p_{\alpha}$ that simply renormalizes the parallel orientational field, $h_{\parallel},$ and has no physical meaning). Such a $B_{\alpha\beta}p_{\beta}$ term is still possible in the diagonal coupling to the orientational field $h_\alpha$, due to a free-energy
coupling of the form $B_{\alpha\beta}p_{\alpha}p_{\beta}.$
The same argument explains why we neglect possible terms $\sim h_{\alpha}p_{\beta}$ or $\sim\Delta\mu p_{\alpha}p_{\beta}$ in the equation for the strain rate. \ra{In the biological context of multicellular migration, the latter corresponds to active matrix remodeling}. The dissipative term in the strain-rate equation is a relaxation term, written in terms of the elastic stress, $\sigma^{\rm el}_{\alpha\beta}$, and relaxation time, $\tau$. For the strain energy that we consider in the linear stability, $f_B={\rm Tr} \left(\tenB-\ln\tenB\right)$, the relaxation term reduces to $-\left(B_{\alpha\beta}-\delta_{\alpha\beta}\right)/\tau$.

In both the polarization-rate and strain-rate equations, we neglect a possible reactive coupling to the solvent strain-rate, $v^\s_{\alpha\beta}.$ This is because we have already made our choice of convective terms in the definitions of the derivatives $L^\s_t$ and $L^\n_t$ [Eq.~(\ref{SMeqLt})]. We do consider reactive couplings to the relative current, $J_\alpha$, which are allowed by the polar symmetry. These are the permeation-alignment and \ra{permeation-deformation} terms ($\lambda$ and $\xi,$ respectively).

The fluxes $\delta\sigma_{\alpha\beta}$ and $f^{\rm rel}_\alpha$  are given by
\begin{align}
\delta\sigma_{\alpha\beta}^{\left(d\right)} & =2\eta_{s}v_{\alpha\beta}^{s},\nonumber\\
\delta\sigma_{\alpha\beta}^{\left(r\right)} & =\left(1-\phi\right)\zeta\Delta\mu Q_{\alpha\beta},\\
\delta f^{(d)}_\alpha & =\frac{1}{\gamma\phi\left(1-\phi\right)}J_{\alpha},\nonumber\\
\delta f^{(r)}_\alpha & =-\left(\lambda h_{\alpha}+\xi H^B_{\alpha\beta}p_{\beta}+\nu\Delta\mu p_{\alpha}\right).\label{SMeqdmu}
\end{align}
For  $\delta\sigma_{\alpha\beta}$, we consider the dissipative viscous stress, in terms of the solvent viscosity, $\eta_\s$, and an active, reactive stress, $\left(1-\phi\right)\zeta\Delta\mu Q_{\alpha\beta}$, proportional to the solvent concentration. The solvent viscosity also depends on the solvent concentration. However, as the solvent strain rate vanishes in the steady state and appears only as a first-order correction term, this concentration dependence does not play a role in the linear theory. It is not taken into account hereafter. We neglect a possible, dissipative coupling between $\delta\sigma_{\alpha\beta}$ and $J_{\alpha}$ (or, equivalently, between $\delta f_\alpha$ and $v^\s_{\alpha\beta}$). A coupling between these fields is already included in the definition of $\delta f_\alpha$ [Eq.~(\ref{eqSMnewfluxes})]. For the force $\delta f_\alpha$, we consider a dissipative force due to network-solvent friction, written in terms of the mobility $\gamma$. The first two reactive terms are determined from the reciprocal, reactive couplings in the polarization rate and strain rate.  The final reactive term, $\nu\Delta\mu p_{\alpha}$, gives rise to an active force.

These constitutive relations describe the dynamics of the strain and polarization fields, as well as the force-balance equation on the gel, using $\partial_\beta \sigma_{\alpha\beta}=0$. Next, we demonstrate how they can be interpreted as a two fluid model, written in terms of separate force-balance equations for each of the components. Namely, the flux $\delta f_\alpha$, conjugate to the relative current, is related to the relative force, $f^{\rm rel}_\alpha$ between the two components.
\subsection{B. Interpretation in terms of a two-fluid model}
\label{SMssec12}
We consider the different contributions to the vanishing total force acting on the gel, $\partial_\beta\sigma_{\alpha\beta}=0$. It includes the force resulting from the Ericksen stress~\cite{Joanny07},
\begin{align}
\partial_{\beta}\sigma_{\alpha\beta}^{\rm Er} & =-n_{\s}\partial_{\alpha}\mu_{\s}-n_{\n}\partial_{\alpha}\mu_{\n}-h_{\beta}\partial_{\alpha}p_{\beta}-H_{\beta\gamma}^{B}\partial_{\alpha}B_{\beta\gamma}\nonumber\\
&=-\partial_\alpha\delta P-\phi\partial_\alpha\bmu-h_{\beta}\partial_{\alpha}p_{\beta}-H_{\beta\gamma}^{B}\partial_{\alpha}B_{\beta\gamma},
\end{align}
where we have made use of the gel incompressibility $n_\s+n_\n=\rho/m$, with $m$ being the molecular mass (assumed equal for both components), and have denoted $\delta P=\rho\mu_{s}/m.$
For simple solvents and in the absence of elasticity, for which the
Ericksen stress reduces to (minus) the osmotic pressure, $\delta
P=P-\Pi$ is the difference between total pressure
and osmotic pressure. The relative chemical potential is $\bmu=\rho\left(\mu_\n-\mu_\s\right)/m$.

The total force is thus given by
\begin{align}
\label{eqSMtot}
\partial_\beta\sigma_{\alpha\beta} & = \partial_\beta \sigma^{\rm el}_{\alpha\beta}-\phi\partial_\alpha\left(\bmu+\delta P\right)-H^B_{\beta\gamma}\partial_\alpha B_{\beta\gamma}\nonumber\\
&+\partial_\beta\left(\delta\sigma_{\alpha\beta}-h_\alpha p_\beta\right)-\left(1-\phi\right)\partial_\alpha \delta P-h_\beta\partial_\alpha p_\beta\nonumber\\&=0.
\end{align}
The first line in the right-hand-side of the equation above is written in terms of network-dependent physical quantities, and the second line in terms of solvent-dependent ones. They can be interpreted as the force on the network and on the solvent, respectively, not including relative forces between the two components, which do not contribute to the total stress, $\sigma_{\alpha\beta}$.

We define
\ba
f^{\n}_\alpha&=\partial_\beta\sigma^{\rm el}_{\alpha\beta}-\phi\partial_\alpha\bmu-H^B_{\beta\gamma}\partial_\alpha B_{\gamma\beta},\nonumber\\
f^{\s}_\alpha&=\partial_\beta \left(\delta\sigma_{\alpha\beta}-h_\alpha p_\beta\right)-h_\beta\partial_\alpha p_\beta\label{SMeq7}.
\end{align}
The force on the network is $f^\n_\alpha-\phi\,\partial_\alpha\delta P$ and the one on the solvent is $f^\s_\alpha-\left(1-\phi\right)\partial_\alpha\delta P$. Comparing to Eqs.~(\ref{eqSMnewfluxes}) and (\ref{eqSMtot}), We find that,
\ba
\label{SMeqftotfrel}
f^\n_\alpha+f^\s_\alpha-\partial_\alpha\delta P&=0,\nonumber\\
\frac{1}{\phi}f^\n_\alpha-\frac{1}{1-\phi}f^\s_\alpha &=\delta f_\alpha.
\end{align}
A linear combination of these two equations yields
\ba
\label{EqSMforceSN}
f^{\n}_\alpha-\phi\partial_\alpha\delta P &=\phi\left(1-\phi\right)\delta f_\alpha,\nonumber\\
f^{\s}_\alpha-\left(1-\phi\right)\partial_\alpha\delta P&=-\phi\left(1-\phi\right)\delta f_\alpha.
\end{align}
These are the separate force-balance equations for the network and solvent, respectively, in a two-fluid model. It is now possible to identify the relative force between the two components, $\phi\left(1-\phi\right)\delta f_\alpha=f^{\rm rel}_\alpha$. Equation~(\ref{EqSMforceSN}) thus yields Eq.~(5) in the main text.
 This explains the interpretation of the \ra{permeation-deformation} and permeation-alignment couplings, which appeared originally  as phenomenological, polar couplings in Eqs.~(\ref{eqp}) and (\ref{eqB}), as relative forces between network and solvent.

In this derivation of the two-fluid model, we have made use of the fact that $\delta\sigma_{\alpha\beta}$ is a stress that originates only in the solvent, as was chosen in our constitutive relations, $\delta\sigma_{\alpha\beta}=2\eta_\s v^\s_{\alpha\beta}+\left(1-\phi\right)\zeta\Delta\mu Q_{\alpha\beta}$. This choice neglects possible network contributions, proportional to $H^B_{\alpha\beta}$ and $\Delta\mu B_{\alpha\beta}$. A two-fluid model can be similarly derived when such network contributions are taken into account, as we demonstrate now.

We write $\delta\sigma_{\alpha\beta}=\delta\sigma^\s_{\alpha\beta}+\delta\sigma^\n_{\alpha\beta}$, where $\delta\sigma^\s_{\alpha\beta}$ is the solvent contribution and $\delta\sigma^\n_{\alpha\beta}$ is the network contribution of the form $\delta\sigma^\n_{\alpha\beta}=a H^B_{\alpha\beta}+b\Delta\mu B_{\alpha\beta}$. In this case, while Eq.~(\ref{eqSprod2a}) still holds, $\delta f_\alpha$ is related differently to the relative force. Following the same arguments as above, we find that
\begin{align}
\label{eqSMnewfluxesa}
\frac{f^{\rm rel}_\alpha}{\phi\left(1-\phi\right)}&=\frac{1}{\phi}\left[\partial_\beta\left(\delta\sigma^\n_{\alpha\beta}+\sigma^{\rm el}_{\alpha\beta}\right)-\phi\partial_{\alpha}\bmu-H^B_{\beta\gamma}\partial_\alpha B_{\beta\gamma}\right]-\frac{1}{1-\phi}\left[\partial_\beta \left(\delta\sigma^\s_{\alpha\beta}-h_\alpha p_\beta\right)-h_\beta \partial_\alpha p_\beta\right].
\end{align}
This yields $\phi\left(1-\phi\right)\delta f_\alpha=f^{\rm rel}_\alpha-\partial_\beta \delta\sigma^\n_{\alpha\beta}$. Inserting this expression in Eq.~(\ref{eqSprod2a}) yields
\begin{align}
\label{eqSprod3b}
\dot{F}&=-\int \D\boldsymbol{r}\left[v^\s_{\alpha\beta}\left(\delta\sigma^\s_{\alpha\beta}+\delta\sigma^\n_{\alpha\beta}\right)+h_{\alpha}\Ls\, p_{\alpha}+H^B_{\alpha\beta}\Ln B_{\alpha\beta}+\frac{J_\alpha}{\phi\left(1-\phi\right)}\left(f^{\rm rel}_\alpha-\partial_\beta\delta\sigma^\n_{\alpha\beta}\right)+\Delta\mu r \right].
\end{align}
This ensures that any choice of $\delta\sigma^\n_{\alpha\beta}$ will be incorporated in two constitutive relations, one conjugate to $v^\s_{\alpha\beta}$ and one conjugate to $J_\alpha$. In this way, the reciprocal terms will depend consistently on the network strain rate, $v^\n_{\alpha\beta}$. The resulting two-fluid model is described by Eq. (\ref{EqSMforceSN}), with the forces,
\ba
f^{\n}_\alpha&=\partial_\beta\left(\delta\sigma^\n_{\alpha\beta}+\sigma^{\rm el}_{\alpha\beta}\right)-\phi\partial_\alpha\bmu-H^B_{\beta\gamma}\partial_\alpha B_{\gamma\beta},\nonumber\\
f^{\s}_\alpha&=\partial_\beta \left(\delta\sigma^\s_{\alpha\beta}-h_\alpha p_\beta\right)-h_\beta\partial_\alpha p_\beta.
\end{align}

\section{\Romannum{2}. Linearized version of the equations}
\label{SMsec2}
In this section we derive the linearized version of the equations,
which is used for the linear stability analysis. We first write the
equations in full form, including explicit expressions for the fields
that are derived from the free energy. Then, we solve the steady-state
equations, and linearize around the steady-state solutions.

\subsection{A. Full form}
\label{SMssec21}
We consider the free energy of Eq. (1) in the Letter,
\begin{align}
F & =\int dr\,\left[f_{\phi}+\frac{1}{2}G\phi\text{Tr}\left(\tenB-\ln \tenB\right)+\left(1-\phi\right)^{2}\left(\frac{1}{2}K\left(\nabla \vecp\right)^{2}+K_{d}\nabla\cdot \vecp\right)-\frac{1}{2}h_{\parallel}\vecp^{2}\right].
\end{align}
Here, we have inserted the Flory Gaussian-chain free-energy density,
$f_{B}=G\phi\text{Tr}\left(\tenB-\ln \tenB\right)/2$ and have neglected the
possible strain-polarization coupling, $f_{Bp}=0.$ The resulting
solvent orientational field is
\begin{align}
h_{\alpha} & =-\frac{\delta F}{\delta p_{\alpha}}=\partial_{\beta}\left[\left(1-\phi\right)^{2}K\partial_{\beta}p_{\alpha}\right]-2K_{d}\left(1-\phi\right)\partial_{\alpha}\phi+h_{\parallel}p_{\alpha}.\label{SMeqh}
\end{align}
Note that the polar-splay contribution to the orientational field is via
the concentration gradient. Otherwise, this term can be integrated
in the free energy to a boundary term (divergence theorem), which
does not contribute to the orientational field. The network molecular
field is given by
\begin{align}
H^B_{\alpha\beta} & =-\frac{\delta F}{\delta B_{\alpha\beta}}=-\frac{1}{2}G\phi\left(\delta_{\alpha\beta}-B_{\alpha\beta}^{-1}\right),
\end{align}
and the elastic stress is
\begin{align}
\sigma^{\rm el}_{\alpha\beta}=-2H^B_{\alpha\gamma}B_{\gamma\beta} & =G\phi\left(B_{\alpha\beta}-\delta_{\alpha\beta}\right).
\end{align}
The relative chemical potential reads
\begin{align}
\bmu & =\frac{\delta F}{\delta\phi}=\bmu_{0}\left(\phi\right)+\frac{1}{2}G\text{Tr}\left(\tenB-\ln \tenB\right)-2\left(1-\phi\right)\left(\frac{1}{2}K\left(\nabla \vecp\right)^{2}+K_{d}\nabla\cdot \vecp\right),\label{SMeqmu}
\end{align}
where $\bmu_{0}$ is derived from the mixing term, $\bmu_{0}=\delta f_{\phi}/\delta\phi-\partial_{\alpha}\delta f_{\phi}/\delta\partial_{\alpha}\phi.$
Equations (\ref{SMeqh})-(\ref{SMeqmu}) relate the fields that appear in the forces and fluxes
of the entropy production to the dynamic fields, $\phi,$ $p_{\alpha}$
and $B_{\alpha\beta},$ whose dynamics we analyze to linear order.

The concentration fields satisfies the continuity equation,
\begin{align}
\partial_{t}\phi+\partial_{\alpha}\left(\phi v_{\alpha}^{n}\right) & =0.
\end{align}
The polarization and strain fields evolve according to Eqs. (2) and (3) in the main text of the Letter, and the velocities are related to the polarization, strain and concentration from the force-balance equations on the solvent and network [Eqs. (4)-(7) in the main text].

\subsection{B. Steady state }
\label{SMssec22}
We search for a steady-state with homogeneous fields $\phi=\phi_{0},$
$p_{\alpha}=p_{\alpha}^{0}=\delta_{\alpha1},$ $B_{\alpha\beta}=B_{\alpha\beta}^{0},$ a
zero COM velocity, $v_{\alpha}=0,$ and a homogeneous relative current,
$J_{\alpha}^{0}=J_{0}p_{\alpha}^{0}.$ As the system is homogeneous,
all the gradient terms vanish, including the convective terms, and
forces acting on the network and solvent. The relative current is
determined from $f_{\alpha}^{rel}=0.$

The relative force depends on the parallel solvent orientational field,
$h_{\parallel}.$ This field is a Lagrange multiplier that ensures $p^2=1.$
Its value is determined by projecting the polarization rate equation
on the polarization. We find that $h_{\parallel}=-\gamma_{1}\lambda J_{0}.$
In addition, the steady-state strain is given by $B_{\alpha\beta}^{0}=\delta_{\alpha\beta}+\xi\tau J_{0}p_{\alpha}^{0}p_{\beta}^{0}.$
As $B_{\alpha\beta}$ is positive definite, this steady state is
possible only for $\xi\tau J_{0}>-1.$ The resulting molecular field
is $H^{B\,(0)}_{\alpha\beta}=-G\phi_{0}\xi\tau J_{0}\left(1+\xi\tau J_{0}\right)^{-1}p_{\alpha}^{0}p_{\beta}^{0}/2.$

The equation $f_{\alpha}^{rel}p_{\alpha}^{0}=0$ then reduces to
\begin{align}
\frac{1}{\gamma}J_{0}-\phi_{0}\left(1-\phi_{0}\right)\left(-\lambda^{2}\gamma_{1}J_{0}-\frac{1}{2}G\phi_{0}\xi^{2}\frac{\tau J_{0}}{1+\xi\tau J_{0}}+\nu\Delta\mu\right) & =0.
\end{align}
This is generally a quadratic equation in $J_{0}.$ As our framework is formulated to linear order in $J_0$, we consider small $\xi\tau J_{0}$ and
retain only linear terms. This yields the relative current,
\begin{align}
J_{0} & =\tilde{\gamma}\phi_{0}\left(1-\phi_{0}\right)\nu\Delta\mu,
\end{align}
 with the renormalized mobility
\begin{align}
\tilde{\gamma} & =\frac{\gamma}{1+\gamma\phi_{0}\left(1-\phi_{0}\right)\left(\lambda^{2}\gamma_{1}+\frac{1}{2}\eta_{n}\xi^{2}\right)},
\end{align}
where $\eta_{n}=G\phi_{0}\tau$ is the network viscosity. The mobility
is effectively decreased, due to the new, polar, relative forces between
the solvent and network. For simplicity, we consider in our work only
linear terms in the new coupling terms, and neglect, $\gamma\phi_{0}\left(1-\phi_{0}\right)\left(\lambda^{2}\gamma_{1}+\frac{1}{2}\eta_{n}\xi^{2}\right)\ll1.$
In this case, the mobility retains its original value, $\tilde{\gamma}\simeq\gamma$
and $J_{0}=\gamma\phi_{0}\left(1-\phi_{0}\right)\nu\Delta\mu.$ This
approximation is re-examined in Sec.~\Romannum{5} of the SM.

\subsection{C. Linearized equations}
\label{SMssec23}
The stability is studied by introducing a small perturbation in the fields at point $\vecr$ and time $t
$ with a wave vector $\vecq=q\hat{y}$ and growth rate $s$. For simplicity, we focus on a 2-dimensional system,
\begin{align}
\phi & =\phi_{0}+\phi^{1}\exp\left(iqy+st\right),\nonumber\\
p_{\alpha} & =\delta_{\alpha1}+p^{1}\delta_{\alpha2}\exp\left(iqy+st\right),\nonumber\\
B_{xx} & =1+\xi\tau J_{0}+B_{xx}^{1}\exp\left(iqy+st\right),\nonumber\\
B_{xy} & =B_{xy}^{1}\exp\left(iqy+st\right),\nonumber\\
B_{yy} & =1+B_{yy}^{1}\exp\left(iqy+st\right).
\end{align}
In particular, the correction to the polarization is in the $y$-direction to preserve $p^2=1$ to linear order.
The amplitudes (denoted with superscript $1$) are assumed small,
and the equations are linearized in these amplitudes. To this linear
order, we consider the following form of the mixing chemical potential,
\begin{align}
\bmu_{0}^{1} & =\chi^{-1}\left(1+l_{\phi}^{2}q^{2}\right)\phi^{1},
\end{align}
where $\chi^{-1}=\delta^{2}f_{\phi}/\delta\phi^{2}$ is the inverse
osmotic compressibility, and $l_{\phi}$ is a Ginzburg-Landau type
correlation length, which suppresses large-$q$ concentration
fluctuations.

The COM velocity and relative current are similarly expanded to linear
order, $v_{\alpha}=v_{\alpha}^{1}\exp\left(iqy+st\right)$ and $J_{\alpha}=J_{0}p_{\alpha}^{0}+J_{\alpha}^{1}\exp\left(iqy+st\right),$
together with the pressure difference, $\delta P=\delta P^{1}\exp\left(iqy+st\right),$
where the reference pressure was conveniently taken to be zero. Incompressibility
yields $v_{2}^{1}=0.$ The pressure can then be determined from the
$y$-component of the force-balance equation on the entire gel,
\begin{align}
\delta P^{1} & =-2i\eta_{s}q\frac{J_{y}^1}{1-\phi_{0}}+\half\zeta\Delta\mu\phi^1+\phi_{0}GB^1_{yy}-\phi_{0}\left[-2iK_{d}\left(1-\phi_{0}\right)qp^{1}+\frac{1+l_{\phi}^{2}q^{2}}{\chi}\phi^{1}\right].
\end{align}
The first term on the RHS originates from the solvent viscosity,
the second term is the compressional, active stress in the $y$-direction due to concentration variations, the third term is the compressional elastic stress in the $y$-direction,
and the last two terms are components of the osmotic contribution
to the pressure.

The components for the relative current can be found from the network
force-balance equation. The $x$-component is simpler, because it
does not involve pressure terms. We find that
\begin{align}
\gamma^{-1}J_{x}^{1} & =iqG\phi_{0}B^1_{xy}+\nu\Delta\mu\left(1-2\phi_{0}\right)\phi^{1}-\frac{1}{2}\xi G\phi_{0}^{2}\left(1-\phi_{0}\right)B^1_{xx}.
\end{align}
The first term on the RHS is the force exerted by the shear strain. The
second term is the correction to the $x$-component of the active,
relative force, due to concentration fluctuations. The third term
is the force due to the \ra{permeation-deformation} mechanism. The permeation-alignment
mechanism vanishes to linear order in $\lambda.$

The $y$-component of the relative current includes the contribution
coming from the pressure. We find that
\begin{align}
\frac{1}{\gamma\phi_{0}\left(1-\phi_{0}\right)}\left[1+2\gamma\eta_{s}\frac{\phi_{0}}{1-\phi_{0}}q^{2}\right]J_{y}^{1} & =\lambda\left[-\left(1-\phi_{0}\right)^{2}Kq^{2}p^{1}-2K_{d}\left(1-\phi_{0}\right)iq\phi^{1}\right]-\frac{1}{2}\xi G\phi_{0}B^1_{xy}+\nu\Delta\mu p^{1}\nonumber\\
 & +iq\left(GB^1_{yy}-\left[-2iK_{d}\left(1-\phi_{0}\right)qp^{1}+\frac{1+l_{\phi}^{2}q^{2}}{\chi}\phi^{1}\right]-\frac{1}{2\left(1-\phi_0\right)}\zeta\Delta\mu\phi^1\right),\label{SMeqJ}
\end{align}
The coefficient on the LHS originates from the viscous term in the
pressure. The first line of the RHS are the relative forces, including
the contributions from permeation alignment ($\sim\lambda$), \ra{permeation deformation}
($\sim\xi$) and active force ($\sim\nu\Delta\mu$). The second line
of the RHS includes the divergence of the weighted difference between
the network stress and solvent stress [compare with the expression for $\delta\,f_\alpha$ in Eq.~(\ref{eqSMnewfluxes})].

The $x$-component of the
solvent velocity is found from the $x$-component of the force-balance
equation on the entire gel, which amounts to equating the shear stress
to zero. This yields
\begin{align}
v_{x}^{s (1)} & =\frac{i}{\eta_{s}q}\left[G\phi_{0}B^1_{xy}+\left(1-\phi_0\right)\zeta\Delta\mu p^{1}+\gamma_{1}\lambda J_{0}p^{1}\right],
\end{align}
where the first term on RHS is the elastic shear stress, the second
term is the active shear stress to linear order, and the last term
is the stress associated with polarization rotation to linear order.

Substituting the above results for the velocities, as well as the
expressions for the chemical potential and orientational fields, we obtain
linear equations in terms of only the dynamic fields. These equations
are
\begin{align}
0 & =\left(s+\frac{q^{2}}{1+l_{\eta}^{2}q^{2}}\left(\frac{l^{2}}{\tau_{\phi}}\left(1+l^{2}q^{2}\right)+\frac{l_\eta^2}{4\tau_a}\right)+\frac{2\lambda l_{p}l_{\phi p}^{2}q^{2}}{\tau_{p}\left(1-\phi_{0}\right)\left(1+l_{\eta}^{2}q^{2}\right)}\right)\phi^{1}\nonumber\\
 &+\left(\frac{iJ_{0}q}{1+l_{\eta}^{2}q^{2}}-\frac{2il_{p}l_{\phi p}^{2}q^{3}}{\tau_{p}\left(1-\phi_{0}\right)\left(1+l_{\eta}^{2}q^{2}\right)}-\frac{i\lambda l_{p}^{2}l_{\phi p}^{2}q^{3}}{\tau_{p}\left(1+l_{\eta}^{2}q^{2}\right)}\right)p^{1} -\frac{\left(1-\phi_{0}\right)^{2}\phi_{0}l_{B}^{2}}{\tau\left(1+l_{\eta}^{2}q^{2}\right)}\left(i\frac{\phi_{0}}{2}\xi qB_{xy}^{1}+q^{2}B_{yy}^{1}\right),\nonumber\\
0 & =\left(i\lambda \frac{q}{1+l_{\eta}^{2}q^{2}}\left(\frac{l^{2}}{\tau_{\phi}}\left(1+l^{2}q^{2}\right)+\frac{l_\eta^2}{4\tau_a}\right)+\frac{2il_{p}q}{\tau_{p}\left(1-\phi_{0}\right)}\right)\phi^{1}\nonumber\\&+\left(s+\frac{l_{p}^{2}q^{2}}{\tau_{p}}+\frac{\lambda J_{0}l_{\eta}^{2}q^{2}}{1+l_{\eta}^{2}q^{2}}+\frac{2\lambda l_{p}l_{\phi p}^{2}q^{2}}{\tau_{p}\left(1-\phi_{0}\right)\left(1+l_{\eta}^{2}q^{2}\right)}\right)p^{1}-i\lambda\frac{\left(1-\phi_{0}\right)^{2}\phi_{0}l_{B}^{2}}{\tau\left(1+l_{\eta}^{2}q^{2}\right)}qB_{yy}^{1},\nonumber\\
0 & =-\xi\frac{1-2\phi_{0}}{\phi_{0}\left(1-\phi_{0}\right)}J_{0}\phi^{1}+\left(s+\frac{1}{\tau}\right)B_{xx}^{1}-\frac{i\phi_{0}\left(1-\phi_{0}\right)\xi l_{B}^{2}q}{\tau}B_{xy}^{1},\nonumber\\
0 & =\frac{i}{2}\xi \frac{q}{1+l_{\eta}^{2}q^{2}}\left(\frac{l^{2}}{\tau_{\phi}}\left(1+l^{2}q^{2}\right)+\frac{l_\eta^2}{4\tau_a}\right)\phi^{1}+\left(\frac{2\lambda l_{\phi p}^{2}J_{0}}{l_{\eta}^{2}\left(1-\phi_{0}\right)^{2}}-\frac{\xi J_{0}}{2}\frac{2+l_{\eta}^{2}q^{2}}{1+l_{\eta}^{2}q^{2}}+\frac{\xi l_{p}l_{\phi p}^{2}q^{2}}{\tau_{p}\left(1-\phi_{0}\right)\left(1+l_{\eta}^{2}q^{2}\right)}+\frac{1}{\tau_{a}}\right)p^{1}\nonumber\\
 & +\left[s+\frac{1}{\tau}\left(1+2\phi_{0}^{2}\frac{l_{B}^{2}}{l_{\eta}^{2}}\right)+\frac{l_{B}^{2}q^{2}}{\tau}\right]B_{xy}^{1}+i\frac{l_{B}^{2}}{2\tau}\xi q\phi_{0}\left(1-\phi_{0}\right)\left(B_{xx}^{1}-\frac{1-\phi_{0}}{1+l_{\eta}^{2}q^{2}}B_{yy}^{1}\right),\nonumber\\
0 & =-\left(\frac{4\lambda l_{p}l_{\phi p}^{2}q^{2}}{\phi_{0}\left(1-\phi_{0}\right)\tau_{p}\left(1+l_{\eta}^{2}q^{2}\right)}+\frac{2}{\phi_{0}}\frac{q^{2}}{1+l_{\eta}^{2}q^{2}}\left(\frac{l^{2}}{\tau_{\phi}}\left(1+l^{2}q^{2}\right)+\frac{l_\eta^2}{4\tau_a}\right)\right)\phi^{1}\nonumber\\&+\left(\frac{2i\lambda l_{p}^{2}l_{\phi p}^{2}q^{3}}{\tau_{p}\phi_{0}\left(1+l_{\eta}^{2}q^{2}\right)}-\frac{2iJ_{0}q}{\phi_{0}\left(1+l_{\eta}^{2}q^{2}\right)}+\frac{4il_{p}l_{\phi p}^{2}q^{3}}{\tau_{p}\phi_{0}\left(1-\phi_{0}\right)\left(1+l_{\eta}^{2}q^{2}\right)}\right)p^{1}\nonumber\\
 & +\frac{i\xi l_{B}^{2}q}{\tau\left(1+l_{\eta}^{2}q^{2}\right)}\phi_{0}\left(1-\phi_{0}\right)^{2}B_{xy}^{1}+\left[s+\frac{1}{\tau}+\frac{2l_{B}^{2}q^{2}}{\tau}\frac{\left(1-\phi_{0}\right)^{2}}{1+l_{\eta}^{2}q^{2}}\right]B_{yy}^{1}.
\end{align}
They describe the time evolution of the fields $\phi^1$, $p^1$, $B_{xx}^1$, $B_{xy}^1$, and $B_{yy}^1$, respectively.
For the sake of brevity, these equations are written in terms of
several lengthscales and timescales that are defined below in Sec.~\Romannum{3}. The equations can be written in matrix form as, $\boldsymbol{M}\cdot\boldsymbol{x}=0$,
where $\boldsymbol{x}=\left(\phi^{1},p^{1},B_{xx}^{1},B_{xy}^{1},B_{yy}^{1}\right)^{T}$
is a vector of the perturbed fields, and $\boldsymbol{M}$ is
the dynamic matrix. \ra{Within the framework of linear stability,} the dispersion relations $s(\vecq)$
are obtained by solving ${\rm det} \tenM=0$, which is a fifth-order polynomial in $s$. The system is stable if $\re\, s<0$ for all the eigenvalues.

\section{\Romannum{3}. Parameters of the theory and non-dimensionalization}
\label{SMsec3}
Our theory includes $14$ parameters: $\phi_{0},$ $l_\phi,$ $\chi,$ $G,$
$K_{d},$ $K,$ $\tau,$ $\gamma,$ $\gamma_{1},$ $\eta_{s},$ $\xi,$
$\lambda,$ $\nu\Delta\mu,$ and $\zeta\Delta\mu.$ These parameters
have units that combine length, time, and energy. Therefore, from
the Buckingham-Pi theorem, there are $11$ independent, dimensionless
parameters. It is convenient to introduce the active, relative current,
$J_{0}=\gamma\phi_{0}\left(1-\phi_{0}\right)\nu\Delta\mu,$ and $12$
parameters in units of length and time.

These length scales are $l_{\phi},$ $l_{p}=K/K_{d},$ $1/\xi,$ $1/\lambda,$
$l_{B}=\sqrt{D_{B}\tau}=\sqrt{G\gamma\tau/\left(1-\phi_{0}\right)},$
$l_{\phi p}=\sqrt{\gamma_{1}\gamma\phi_{0}\left(1-\phi_{0}\right)}$
and $l_{\eta}=\sqrt{2\gamma\eta_{s}\phi_{0}/\left(1-\phi_{0}\right)}.$
$l_{\eta}$ is proportional to the length over which dissipation due
to network-solvent friction matches the dissipation due to the solvent
viscosity. \ra{For poroelastic materials, $l_\eta$ is proportional to the mesh size.} $l_{\phi p}$ is proportional to the length over dissipation due
to network-solvent friction matches the dissipation due to the polar rotational viscosity.

The time scales are $\tau,$ $\tau_{p}=l_{p}^{2}/D_{p}=K\gamma_{1}/\left[K_{d}\left(1-\phi_{0}\right)\right]^{2}$,
$\tau_{\phi}=l_{\phi}^{2}/D_{\phi}=\chi l_{\phi}^{2}/\left[\gamma\phi_{0}\left(1-\phi_{0}\right)\right]$,
and $\tau_{a}=\eta_{s}/\left[\left(1-\phi_0\right)\zeta\Delta\mu\right].$ The latter is
an active time scale, that describes the rate in which the solvent
need to be sheared, so that the viscous stress matches the active
stress.

Among the above parameters, some can be negative. These are $l_{p},$
$\xi,$ $\lambda,$ $J_{0}$ (polar terms), and $\tau_{a}$. It is
possible to rescale the lengths by $l_\phi$ and times by $\tau_{\phi}.$
This yields, together with $\phi_{0}$, the $11$ desired
dimensionless parameters.

The large number of parameters makes the analysis of the system challenging.
This is why we have made several approximations (see Sec.~\Romannum{5}) and have focused on novel permeation
instabilities that arise to linear order in the polar terms $J_{0},$
$\xi,$ and $\lambda.$ Further analysis beyond
the scope of our calculation is reserved for future studies. Furthermore, the 14 above-mentioned quantities are already a reduced
number of system parameters. Other relevant quantities are, for example:
additional elastic moduli (corresponding to other Poisson's ratios
and possible nonlinearities), bulk viscoelastic relaxation time and
bulk viscosity (taken here as equal to the corresponding shear values),
and configuration-dependent friction coefficients (e.g., $\gamma_{\alpha\beta}=\gamma\delta_{\alpha\beta}+\gamma_{Q}Q_{\alpha\beta}$).
We did not retain such quantities for the purposes of our generic,
physical theory. They may be relevant for a quantitative
analysis of experiments.

\section{\Romannum{4}. Large-wavenumber stability}
\label{SMsec4}
While our hydrodynamic theory is valid for small $q$ values, we test
whether our dynamic variables are stable for large wavenumbers. We
expand the coefficients of the determinant, $\text{det}\boldsymbol{M}$,
to highest order in $q$, and find that
\begin{align}
s^{5}+a_{2}q^{2}s^{4}+a_{4}q^{4}s^{3}+a_{6}q^{6}\left(s^{2}+2\frac{s}{\tau}+\frac{1}{\tau^{2}}\right) & =0,
\end{align}
 with the coefficients
\begin{align}
a_{2} & =\left(\frac{l_{\phi}^{4}}{l_{\eta}^{2}\tau_{\phi}}+\frac{l_{p}^{2}}{\tau_{p}}+\frac{l_{B}^{2}}{\tau}\frac{1}{1+\lambda^{2}l_{\phi p}^{2}}\right),\nonumber\\
a_{4} & =\frac{l_{\phi}^{4}}{l_{\eta}^{2}\tau_{\phi}}\frac{l_{p}^{2}}{\tau_{p}}+\left(\frac{l_{\phi}^{4}}{l_{\eta}^{2}\tau_{\phi}}+\frac{l_{p}^{2}}{\tau_{p}}\right)\frac{l_{B}^{2}}{\tau}\frac{1}{1+\lambda^{2}l_{\phi p}^{2}},\nonumber\\
a_{6} & =\frac{1}{\tau\tau_{p}\tau_{\phi}}\frac{1}{1+\lambda^{2}l_{\phi p}^{2}}\frac{l_{\phi}^{4}l_{p}^{2}l_{B}^{2}}{l_{\eta}^{2}}.
\end{align}

There are two possible types of solutions. First, a relaxation solution
$s=s_{0}.$ In this case, only the parenthesis
multiplying $q^{6}$ is required to vanish. These are the elastic
relaxations, $s_{1,2}=-1/\tau$ of the two compressional strains, $B_{xx}$ and $B_{yy}$. This result is related to the fact that we consider osmotic diffusion as the sole origin of strain diffusion. In this limit, $B_{xx}$ has a vanishing diffusion coefficient for wave vectors in the $y$-direction. The diffusion coefficient of $B_{yy}$, on the other hand, decays to zero, because the effective mobility coefficient, $\gamma$, in the $y$-direction scales as $1/q^2$ in this limit, due to solvent compressibility (see Sec.~\Romannum{6}).

Another type of solution is diffusion, $s=-Dq^{2}.$ Inserting this solution in the equation, we find that the highest-order terms scale as $q^{10}$
and $D$ solves the equation $D^{3}-a_{2}D^{2}+a_{4}D-a_{6}=0.$ The
three solutions are given by $D=l_{p}^{2}/\tau_{p}$, corresponding
to rotational diffusion, $D=l_{B}^{2}/\left[\left(1+\lambda^{2}l_{\phi p}^{2}\right)\tau\right]$,
corresponding to shear strain diffusion, and $D=l_{\phi}^{4}/\left(l_{\eta}^{2}\tau_{\phi}\right).$ The latter diffusion constant is related to the concentration $\phi^1$, but is different than the small-$q$
diffusion coefficient, $l^2_\phi/\tau_\phi$. The diffusion in this case is not osmotic, but describes mass-conserving compression/dilations in the $y$-direction. The local concentration is then given by $\phi_0+\phi^1=1/v'\simeq \phi_0\left(1-B^1_{yy}/2\right)$, where $v'$ is the volume of a network element after a $B^1_{yy}$ deformation.

All the growth rates are negative, meaning that
the system is stable for large $q$ values. We emphasize that the theory is hydrodynamic. It was derived from a gradient expansion, and is adequate for small wave vectors. Terms that we have neglected may become important in the large-$q$ limit. This Section serves only to demonstrate that our theory is consistent and does not result in large-$q$ instabilities. We make no further predictions in this limit.

\section{\Romannum{5}. Validity }
\label{SMsec5}
The theory depends, as was explained in Sec.~\Romannum{3}, on 11 dimensionless
parameters. For the sake of simplicity, we perform the analysis to
linear order in the \ra{permeation-deformation} and permeation-alignment
parameters ($\xi$ and $\lambda$, respectively). This implies a lengthscale
$l,$ such that $\xi l$ and $\lambda l$ are considered to be small.
For example, the steady-state relative current, $J_{0}=\gamma\phi_{0}\left(1-\phi_{0}\right)\nu\Delta\mu$
was derived assuming that $\gamma\phi_{0}\left(1-\phi_{0}\right)\left(\lambda^{2}\gamma_{1}+\frac{1}{2}\eta_{n}\xi^{2}\right)\ll1.$
In terms of the lengthscales that we have introduced in Sec.~\Romannum{3} this condition
can be written as
\begin{align}
\left(\lambda l_{\phi p}\right)^{2}+\frac{1}{2}\left[\phi_{0}\left(1-\phi_{0}\right)\xi l_{B}\right]^{2} & \ll1.
\end{align}
Below we review similar conditions that were used as part of our calculation.

\subsection{A. Permeation deformation $\left(\xi\protect\ne0\right)$}
\label{SMssec51}
The eigenvector associated with the concentration and polarization
variables, which is a linear combination of them for $q=0$, has
a dispersion relation $s=iuq-Dq^{2}.$ The velocity, $u,$ to quadratic
order in $\xi,$ is given by
\begin{align}
u & =\sqrt{\frac{2}{1-\phi_{0}}\frac{l_{p}}{\tau_{p}}\left(J_0+J_{\xi}\left[1-J_{0}\xi\left(\tau+\tau_{a}\right)\right]\right)},
\end{align}
where
\begin{align}
J_{\xi} & =\frac{1}{2}\frac{\xi}{\tau_{a}}\left[\phi_{0}\left(1-\phi_{0}\right)\right]^{2}\frac{l_{B}^{2}}{1+2\phi^{2}l_{B}^{2}/l_{\eta}^{2}}.
\end{align}
Therefore, we have considered the limit where $\left|J_{0}\xi\left(\tau+\tau_{a}\right)\right|\ll1.$
The diffusion coefficient, $D,$ to quadratic order in $\xi,$ reads
\begin{align}
D & =\frac{1}{2}\left(D_{p}+D_{\phi}+D_\zeta\right)+\frac{1}{2}D_{\xi}\left[1+\xi\left(\frac{1-\phi_{0}}{4}\frac{\tau_{a}}{\tau}\frac{\tau_{p}}{l_p}\left(\frac{l_\phi^2}{\tau_\phi}+\frac{l_\eta^2}{4\tau_a}\right)\left(1+2\phi_{0}^{2}\frac{l_{B}^{2}}{l_{\eta}^{2}}\right)-J_{0}\left(\tau+\tau_{a}\right)\right)\right].
\end{align}
Therefore, in addition to our previous condition, we require here
that
\begin{align}
\left|\xi \frac{\tau_{a}}{\tau}\frac{\tau_{p}}{l_p}\left(\frac{l_\phi^2}{\tau_\phi}+\frac{l_\eta^2}{4\tau_a}\right)\right|&\ll1.
\end{align}
The eigenvector associated with the shear-strain variable has a dispersion
relation $s=s_{0}-Dq^{2}.$ The relaxation time, up to quadratic order
in $\xi$ is given by
\begin{align}
s_{0} & =-\frac{1}{\tau}\left[1+\frac{\eta_{n}}{\eta_{s}}+\left(\frac{1}{2}\phi_{0}\left(1-\phi_{0}\right)l_{B}\xi\right)^{2}\right].
\end{align}
We have thus considered $\phi_{0}\left(1-\phi_{0}\right)l_{B}\xi\ll1$
(as was required for the steady-state relative current). Note, however,
that this new term demonstrates a new possible mechanism for elastic
relaxation through a feedback between the shear strain and relative
current.

The diffusion coefficient, up to quadratic order in $\xi,$ is
\begin{align}
D & =D_{B}-D_{\xi}\left(1-J_{0}\xi\left(\tau+\tau_{a}\right)+\xi\frac{\tau_{p}}{l_{p}}\frac{\tau_{a}}{\tau^{2}}\left(1+2\phi_{0}^{2}\frac{l_{B}^{2}}{l_{\eta}^{2}}\right)\left[\frac{1}{2}\left(\frac{1-\phi_{0}}{\phi_{0}}\right)^{2}l_{\eta}^{2}\left(1+2\phi_{0}\frac{l_{B}^{2}}{l_{\eta}^{2}}\right)+\left(1-\phi_{0}\right)\frac{\tau}{4}\left(\frac{l_{\phi}^{2}}{\tau_\phi}+\frac{l_\eta^2}{4\tau_a}\right)\right]\right).
\end{align}
Therefore, in addition to our previous conditions, we consider here
that
\begin{align}
\frac{1}{2}\left(\frac{1-\phi_{0}}{\phi_{0}}\right)^{2}\left|\frac{\xi l_{\eta}^{2}}{l_{p}}\frac{\tau_{a}\tau_{p}}{\tau^{2}}\right| & \ll1.
\end{align}

The permeation instabilities that we report in the Letter are allowed
by these conditions. They can be satisfied with sufficiently
large active stresses, corresponding to small $\tau_{a}$ and large
$J_{\xi}$ and $D_{\xi}$ in absolute values.

\subsection{B. Permeation alignment $\left(\lambda\protect\ne0\right)$}
\label{SMssec52}
We examine the corrections to our results in the presence of the permeation-alignment
coupling, due to higher-order terms in the parameter, $\lambda.$
We find that the eigenvector associated with the polarization has
a constant growth rate ($\sim q^0$), which is cubic in $\lambda$, $s_{0}=-\lambda^{3}l_{\phi p}^{2}J_{0}.$ In particular, this leads to an instability for $\lambda J_{0}<0.$ We consider this term to be negligible, $s_{0}\tau\ll1.$ In this case, as is presented in the main text, the eigenvector associated with
polarization and concentration has a growth rate $s=iuq-Dq^{2}.$
The velocity, $u,$ is unchanged in this limit. The diffusion coefficient,
however, has a correction
\begin{align}
D & =\frac{1}{2}\left[\left(1+\frac{1}{2}\lambda^{2}l_{\phi p}^{2}\right)D_{p}+D_{\phi}+D_\zeta+D_{\lambda}\right].
\end{align}
This correction increases the effective angular diffusion, and can
only be stabilizing. It is negligible for $\lambda l_{\phi p}\ll1$
(as was required for the steady-state relative current).

This condition also allows for finite $\lambda J_{0}\tau$, as is
required for the possible strain instability in the main text, with negligible
$s_{0}\tau.$ The polarization-concentration instability is also allowed
by this condition. The velocity $u$ can become imaginary depending
on the sign of $l_{p}J_{0}$ and $\lambda J_{0},$ and the diffusion
coefficient can become negative, for example, for sufficiently large
elastic modulus, $G\chi\gg1,$ such that $D_{\lambda}$ is large compared
to $D_{\phi}$ in absolute value. Finally, the diffusion coefficient
of $B_{yy}$ has a correction of order $s_{0}\tau,$ which is negligible
within our limits.

\section{\Romannum{6}. Passive Instability}
\label{SMsec6}
In the main text, we have presented active permeation instabilities that
occur in the presence of new polar coupling terms. It is worth noting
that a passive, polar gel can also become unstable, even in the absence
of the permeation-alignment and \ra{permeation-deformation} mechanisms~\cite{Blankschtein85,Hinshaw88}.
This instability originates from the coupling between the polarization
and concentration. Its eigenmode is a linear combination of the concentration and polarization for $q=0$. 
Below, we derive the criterion for this instability from the dynamic
equations. This derivation allows to analyze
the effects of the new permeation couplings that are allowed even in the absence of activity, $\Delta\mu=0$. We demonstrate that the
instability criterion is unaffected by the new permeation couplings.

\subsection{A. Passive instability for $\lambda=\xi=0$}
\label{SMssec61}
As before, we linearize around the homogeneous steady state in the
passive case (with $J_{0}=0$), for a wave-vector that is perpendicular
to the original polarization, and write the equations in Fourier space.
First, we ignore the solvent viscosity and network elasticity. The
equations for the polarization and concentration are
\begin{align}
sp^{1} & =\frac{1}{\gamma_{1}}h_{2}=-\frac{1}{\gamma_{1}}\left[\left(1-\phi_{0}\right)^{2}Kq^{2}p^{1}+2K_{d}\left(1-\phi_{0}\right)iq\phi^{1}\right],\nonumber\\
s\phi^{1} & =-iqJ_{y}^1=-\gamma\phi_{0}\left(1-\phi_{0}\right)q^{2}\left[\chi^{-1}\left(1+l_{\phi}^{2}q^{2}\right)\phi^{1}-2i\left(1-\phi_{0}\right)K_{d}qp^{1}\right].
\end{align}
In terms of the lengthscales and time scales that we have introduced
above, these equations are given by
\begin{align}
sp^{1} & =-\frac{1}{\tau_{p}}\left(l_{p}^{2}q^{2}p^{1}+\frac{2i}{1-\phi_{0}}l_{p}q\phi^{1}\right),\nonumber\\
s\phi^{1} & =-q^{2}\left[\frac{l_{\phi}^{2}}{\tau_{\phi}}\left(1+l_{\phi}^{2}q^{2}\right)\phi^{1}-\frac{2i}{1-\phi_{0}}\,l_{\phi p}^{2}\frac{l_{p}}{\tau_{p}}qp^{1}\right].
\end{align}
This set of linear equations has a non-trivial solution when the determinant
of coefficients vanishes,
\begin{align}
s^{2}+\left[\frac{l_{p}^{2}}{\tau_{p}}+\frac{l_{\phi}^{2}}{\tau_{\phi}}\left(1+l_{\phi}^{2}q^{2}\right)\right]q^{2}s+\frac{l_{p}^{2}}{\tau_{p}}\left[\frac{l_{\phi}^{2}}{\tau_{\phi}}\left(1+l_{\phi}^{2}q^{2}\right)-\left(\frac{2}{1-\phi_{0}}\right)^{2}\frac{l_{\phi p}^{2}}{\tau_{p}}\right]q^{4} & =0.
\end{align}
The linear term in $s$ is always positive. The solution has a positive
real part only when the coefficient independent of $s$ is negative. An instability
occurs for
\begin{align}
\left(\frac{2}{1-\phi_{0}}\right)^{2}\frac{l_{\phi p}^{2}}{\tau_{p}} & >\frac{l_{\phi}^{2}}{\tau_{\phi}}\left(1+l_{\phi}^{2}q^{2}\right).
\end{align}
This is equivalent to $K<4K_{d}^{2}\chi/\left(1+l_{\phi}^{2}q^{2}\right).$
This mechanism involves only diagonal transport coefficients and originates
only from the concentration-polarization coupling in the free energy.

We now consider the effects of the solvent viscosity and network elasticity.
Adding the solvent viscosity merely renormalizes the friction according
to $1/\gamma\to\left(1+l_{\eta}^{2}q^{2}\right)/\gamma$ [see Eq.~(\ref{SMeqJ})].
Network elasticity introduces a network force due to the elongation
stress in the $y$-direction, $iqG\phi_{0}B^1_{yy}.$ The strain variable,
because of the convected derivative, evolves as
\begin{align}
\left(s+\frac{1}{\tau}\right)B^1_{yy} & =\frac{2}{\phi_{0}}iqJ_{y}^1.
\end{align}
 The new equation for the current is, therefore,
\begin{align}
\frac{1+\left[l_{\eta}^{2}+2\left(1-\phi_{0}\right)^{2}\frac{l_{B}^{2}}{1+s\tau}\right]q^{2}}{\gamma\phi_{0}\left(1-\phi_{0}\right)}J_{y}^1 & =-iq\left[\frac{l_{\phi}^{2}}{\tau_{\phi}}\left(1+l_{\phi}^{2}q^{2}\right)\phi^{1}-\frac{2i}{1-\phi_{0}}\,l_{\phi p}^{2}\frac{l_{p}}{\tau_{p}}qp^{1}\right]
\end{align}
The elasticity can be considered as an $s$-dependent
correction to the friction coefficient. Note that, assuming an instability
($s>0$), the new friction coefficient remains positive, and our previous
analysis holds with a renormalized $\gamma$. As the instability criterion
is independent of $\gamma,$ it is still given by $K<4K_{d}^{2}\chi/\left(1+l_{\phi}^{2}q^{2}\right).$

\subsection{B. Passive instability with permeation alignment $\left(\lambda\protect\ne0\right)$}
\label{SMssec62}
The permeation alignment induces a relative force between the network
and solvent, which modifies the relative current,
\begin{align}
J_{y}^1 & =-iq\left[\frac{l_{\phi}^{2}}{\tau_{\phi}}\left(1+l_{\phi}^{2}q^{2}\right)\phi^{1}-\frac{2i}{1-\phi_{0}}\,l_{\phi p}^{2}\frac{l_{p}}{\tau_{p}}qp^{1}\right]-\lambda\frac{l_{\phi p}^{2}}{\tau_{p}}\left(l_{p}^{2}q^{2}p^{1}+\frac{2i}{1-\phi_{0}}l_{p}q\phi^{1}\right),
\end{align}
and enters the polarization-rate equation as $sp^{1}=h_{y}^1/\gamma_{1}+\lambda J_{y}^1.$
The equations for the polarization and concentration now read
\begin{align}
sp^{1} & =-\frac{l_{p}^{2}q^{2}}{\tau_{p}}\left[1+\lambda\left(\lambda+\frac{2}{1-\phi_{0}}\frac{1}{l_{p}}\right)l_{\phi p}^{2}\right]p^{1}-\left[\frac{2}{1-\phi_{0}}\left(1+\lambda^{2}l_{\phi p}^{2}\right)+\frac{\lambda l_{\phi}^{2}}{l_{p}}\frac{\tau_{p}}{\tau_{\phi}}\left(1+l_{\phi}^{2}q^{2}\right)\right]i\frac{l_{p}}{\tau_{p}}q\phi^{1},\nonumber\\
s\phi^{1} & =-\frac{l_{\phi}^{2}}{\tau_{\phi}}q^{2}\left[1+l_{\phi}^{2}q^{2}+\frac{2}{1-\phi_{0}}l_{p}\lambda\frac{l_{\phi p}^{2}}{l_{\phi}^{2}}\frac{\tau_{\phi}}{\tau_{p}}\right]\phi^{1}+\left(\frac{2}{1-\phi_{0}}+\lambda l_{p}\right)\,l_{\phi p}^{2}i\frac{l_{p}}{\tau_{p}}q^{3}p^{1}.
\end{align}
A non-trivial solution exists for
\begin{align}
s^{2}+\left[\frac{l_{p}^{2}}{\tau_{p}}+\frac{l_{\phi}^{2}}{\tau_{\phi}}\left(1+l_{\phi}^{2}q^{2}\right)+\lambda\left(\lambda+\frac{4}{1-\phi_{0}}\frac{1}{l_{p}}\right)l_{\phi p}^{2}\frac{l_{p}^{2}}{\tau_{p}}\right]q^{2}s+\frac{l_{p}^{2}}{\tau_{p}}\left[\frac{l_{\phi}^{2}}{\tau_{\phi}}\left(1+l_{\phi}^{2}q^{2}\right)-\left(\frac{2}{1-\phi_{0}}\right)^{2}\frac{l_{\phi p}^{2}}{\tau_{p}}\right]q^{4} & =0.
\end{align}
 The permeation-alignment parameter, $\lambda$, enters the equation only in the coefficient of the term linear in $s$. It can induce a new instability only if the term linear in $s$ is negative. The minimal value of the coefficient is obtained
for $\lambda=-2/\left[\left(1-\phi_{0}\right)l_{p}\right].$ In this
case, the equation reduces to
\begin{align}
s^{2}+\left[\frac{l_{p}^{2}}{\tau_{p}}+\frac{l_{\phi}^{2}}{\tau_{\phi}}\left(1+l_{\phi}^{2}q^{2}\right)-\left(\frac{2}{1-\phi_{0}}\right)^{2}\frac{l_{\phi p}^{2}}{\tau_{p}}\right]q^{2}s+\frac{l_{p}^{2}}{\tau_{p}}\left[\frac{l_{\phi}^{2}}{\tau_{\phi}}\left(1+l_{\phi}^{2}q^{2}\right)-\left(\frac{2}{1-\phi_{0}}\right)^{2}\frac{l_{\phi p}^{2}}{\tau_{p}}\right]q^{4} & =0.
\end{align}
The coefficient of the term linear in $s$ can become negative only if the constant term in $s$ is itself negative. Therefore, the criterion for instability
is not changed. In this calculation we did not treat the viscosity
and elasticity explicitly. They can be absorbed in $\gamma$ (and the
resulting lengthscales), as was explained above.

\subsection{C. Passive instability with permeation deformation $\left(\xi\protect\ne0\right)$}
\label{SMssec63}
\ra{Permeation deformation} induces a relative force between the network
and solvent, which modifies the relative current,
\begin{align}
\left(1+\left[l_{\eta}^{2}+2\left(1-\phi_{0}\right)^{2}\frac{l_{B}^{2}}{1+s\tau}\right]q^{2}\right)J_{y}^1 & =-iq\left[\frac{l_{\phi}^{2}}{\tau_{\phi}}\left(1+l_{\phi}^{2}q^{2}\right)\phi^{1}-\frac{2i}{1-\phi_{0}}\,l_{\phi p}^{2}\frac{l_{p}}{\tau_{p}}qp^{1}\right]-\frac{1}{2}\xi\left[\phi_{0}\left(1-\phi_{0}\right)\right]^{2}\frac{l_{B}^{2}}{\tau}B^1_{xy}.
\end{align}
Here we treat the viscosity and elasticity explicitly. This is required
because the strain evolves differently, due to the \ra{permeation-deformation}
mechanism. The remaining strains evolve as
\begin{align}
\left(1+s\tau\right)B^1_{xy} & =\frac{1}{2}\xi\tau J_{y}^1+iq\tau v_{x}^{n (1)},\nonumber\\
\left(1+s\tau\right)B^1_{xx} & =\xi\tau J_{x}^1.
\end{align}
The relative current in the $x$-direction is induced by the shear
stress and \ra{permeation-deformation} contribution, according to
\begin{align}
J_{x}^1 & =\frac{l_{B}^{2}}{\tau}\phi_{0}\left(1-\phi_{0}\right)\left[iqB^1_{xy}-\frac{1}{2}\phi_{0}\left(1-\phi_{0}\right)\xi B^1_{xx}\right].
\end{align}
The solvent velocity in the $x$-direction is found from the force-balance
on the gel in the $x$-direction. Equating the total shear stress
to zero yields $iq\eta_{s}v_{x}^{s (1)}=-G\phi_{0}B^1_{xy}.$ The network
velocity is found from $v_{x}^{n (1)}=v_{x}^{s (1)}+J_{x}^1/\left[\phi_{0}\left(1-\phi_{0}\right)\right],$ as
\begin{align}
v_{x}^{n (1)} & =\frac{l_{B}^{2}}{\tau}\left[iqB^1_{xy}-\frac{1}{2}\phi_{0}\left(1-\phi_{0}\right)\xi B^1_{xx}\right]+2i\phi^{2}\frac{l_{B}^{2}}{l_{\eta}^{2}}\frac{1}{q\tau}B^1_{xy}.
\end{align}
We can now find the shear strain in terms of the relative current
\begin{align}
B^1_{xy} & =\frac{1}{2}\xi\tau\left(1+s\tau+2\phi^{2}\frac{l_{B}^{2}}{l_{\eta}^{2}}+\frac{1+s\tau}{1+\frac{\tau}{\tau_{\xi}}+s\tau}l_{B}^{2}q^{2}\right)^{-1}J_{y}^1,
\end{align}
where $\tau/\tau_{\xi}=\left[\phi_{0}\left(1-\phi_{0}\right)\xi l_{B}\right]^{2}/2.$
 Inserting this back in the equation for the relative current yields
\begin{align}
J_{y}^1 & =-iq\frac{\frac{l_{\phi}^{2}}{\tau_{\phi}}\left(1+l_{\phi}^{2}q^{2}\right)\phi^{1}-\frac{2i}{1-\phi_{0}}\,l_{\phi p}^{2}\frac{l_{p}}{\tau_{p}}qp^{1}}{1+\left[l_{\eta}^{2}+2\left(1-\phi_{0}\right)^{2}\frac{l_{B}^{2}}{1+s\tau}\right]q^{2}+\frac{1}{2}\frac{\tau}{\tau_{\xi}}\left(1+s\tau+2\phi^{2}\frac{l_{B}^{2}}{l_{\eta}^{2}}+\frac{1+s\tau}{1+\frac{\tau}{\tau_{\xi}}+s\tau}l_{B}^{2}q^{2}\right)^{-1}}.
\end{align}
Therefore, the \ra{permeation-deformation} mechanism results in a renormalized
$s-$ and $q-$dependent friction coefficient. As the new coefficient
is positive for $s>0,$ We can renormalize our lengthscales and timescales,
as we have done above, and obtain the same criterion for instability
as in the $\lambda=\xi=0$ case.

We have assumed in our calculations that $\phi^{1}$ and $p^{1}$
do not vanish. The solution $\phi^{1}=0$ and $p^{1}=0,$ results
in $J_{y}^1=0,$ and therefore $B^1_{xy}=0.$ This infers that $B^1_{xx}$
and $B^1_{yy}$ vanish as well. This shows that our calculation above
holds for any eigenvector of the linear equations.

 \small



\begin{thebibliography}{99}
\bibitem{Lie} The convective terms in Eqs. (\ref{eq2}) and (\ref{eq3}) correspond to Lie derivatives~\cite{holzapfel} \ra{There are other possible choices of convective terms~\cite{SI}.} A similar and more general constitutive equation for one-component, nematic gels is used in Refs.~\cite{Hemingway14,Hemingway16}.
\bibitem{shearinstability} The active shear-alignment instability of Ref.~\cite{voituriez05} does not occur in our case, where the wave-vector is perpendicular to the steady-state polarization and the shear-alignment parameter is $\nu=-1$. The active concentration instability due to director variations of Ref.~\cite{Mishra06} does not occur in our case. Here, the active stress generates, to linear order, a current in the $x$-direction, which does not affect the concentration.
\bibitem{strain} The left Cauchy-Green strain tensor has a simple microscopic interpretation for polymer chains, $\tenB=\langle \vecr\vecr\rangle/l_0^2$, where $\vecr$ is a network segment and $l_0$ is the average segment length in the reference state. A reference state $\langle r_\alpha r_\beta\rangle=l_0^2\delta_{\alpha\beta}$ that is subjected to a deformation tensor $r'_\alpha= E_{\alpha\beta}r_\beta$, deforms as $\langle r'_\alpha r'_\beta\rangle/l_0^2=E_{\alpha\gamma}E_{\beta\gamma}$.
\end{thebibliography}
\end{document}